# Glassy Dynamics in a Molecular Liquid

Shubham Kumar, Sarmistha Sarkar and Biman Bagchi*

*Solid State and Structural Chemistry Unit, Indian Institute of Science, Bangalore 560012, India.*

**\*Email: bbagchi@iisc.ac.in; profbiman@gmail.com**

## Abstract

A universal dynamical crossover temperature, $T_{cr}$, in glassy liquids, associated with the α-β bifurcation temperature, $T_B$, has been observed in dielectric spectroscopy and other experiments. $T_{cr}$ lies significantly above the glass transition temperature. Here, we introduce a new class of glass-forming liquids, binary mixtures of prolate and oblate ellipsoids. This model system exhibits sharp thermodynamic and dynamic anomalies, such as the specific heat jump during heating and a sharp variation in the thermal expansion coefficient around a temperature identified as the glass transition temperature, $T_g$. The same temperature is obtained from the fit of the calculated relaxation times to the Vogel-Fulcher-Tammann (VFT) form. As the temperature is lowered, the single peak rotational relaxation spectrum splits into two at a temperature $T_B$ significantly above the estimated $T_g$. Similar bifurcation is also observed in the distribution of short-to-intermediate time translational diffusion. Interrogation of the two peaks reveals a lower extent of dynamic heterogeneity in the population of the faster mode. We observe an unexpected appearance of a sharp peak in the product of rotational relaxation time $\tau_2$ and diffusion constant $D$ at a temperature $T_{cr}$, close to $T_B$, but above the glass transition temperature. Additionally, we coarse-grain the system into cubic boxes, each containing, on average, ~62 particles, to study the average dynamical properties. Clear evidence of large-scale sudden changes in the diffusion coefficient and rotational correlation time signals first-order transitions between low and high-mobility domains.

## I. INTRODUCTION

Many liquids form disordered, amorphous solids if the temperature is rapidly lowered below their melting point, avoiding crystallization.[1–10] During this cooling process and subsequent heating, the glassy systems exhibit fascinating thermodynamic and dynamic anomalies. Sharp changes in properties are found to occur near a certain temperature, referred to as the glass transition temperature, $T_g$.[1,2,11–13,3–10] When the glass transition is approached from above, dynamical parameters such as the structural relaxation time or the dielectric relaxation time are found to diverge, indicating a large-scale slowdown of relaxation in the supercooled liquids.[6–15] Additionally, the viscosity tends to diverge.[14–16] Understanding the molecular mechanism of the slow dynamics in glass transitions has remained a challenging and central problem for condensed matter science. Fortunately, many excellent reviews and papers exist that have summarized the progress in the area.[1-30]

The well-known bifurcation of the dielectric relaxation spectrum observed in 1967 by Johari and Goldstein, also known as the α-β bifurcation, has continued to draw attention.[31,32] This phenomenon appears to be universal and observed in all different glass-forming liquids, although the underlying molecular origin could be different in different systems.[33–35] It is widely believed that the crossover embodied in Johari-Goldstein (JG) bifurcation holds the key to many of the observed anomalies of the glass transition phenomena that continue to appear below the bifurcation temperature ($T_B$).[36,37] More recently, in an interesting study using quasi-elastic neutron scattering (QENS), Cicerone et al. observed a similar bifurcation phenomenon in a glass-forming liquid but on a much shorter time scale, in the ns to ps time window.[38] The authors performed QENS studies on five liquids, namely, propylene carbonate (PC), propylene glycol (PG), glycerol, ortho-terphenyl (OTP), and sorbitol. The bifurcation observed resembled Johari-Goldstein splitting, although the time scales differ by several orders of magnitude.

It then appears to make sense to search for this bifurcation and to try to understand it through computer simulations. Unfortunately, most of the simulation studies have been restricted to spherical model systems.[6,39–41] The real molecules are hardly spherical. While studies have made notable advancements in elucidating the properties of colloidal glasses,[42–46] the intricate influence of orientation on the dynamics of glassy liquids continues to be a subject of active exploration.[47–56] Ideally, one would like to observe the two bifurcations to emerge simultaneously, as would be predicted by general theories like the random first-order theory (RFOT)[57,58] and the mode coupling theory (MCT).[48,49] We have discussed below that derails can be different because the packing of anisotropic molecules can be different from spherical molecules.

The initial studies of Kauzmann, Gibbs, and DiMarzio gave rise to the entropy crisis theory,[1–3] while Cohen, Turnbull, and Grest [4,5] introduced a theory based on the disappearance (or reappearance) of connected free volume via percolation. As mentioned above, several more sophisticated theories have been developed, like the mode coupling theory (MCT),[39–41,48,49,52] the inherent structure-based analysis,[15,30,59] and the random first-order theory (ROFT).[10,37,57–66] All these theories have thrown considerable light on many aspects of this intriguing problem, yet many questions remain unanswered. Many of these theories find evidence of a *crossover temperature* ($T_{cr}$), which is offered as the explanation of the Johari-Goldstein α-β bifurcation temperature.[37,67] Mode coupling theory predicts a crossover temperature by fitting the time constant of density correlation decay to an anticipated power law.[60,63–66] This crossover temperature is found to lie above the actual glass transition temperature.[67–69] Inherent structure analysis finds a crossover temperature from diffusive dynamics to a landscape-influenced regime when relaxation becomes non-exponential.[59,61,70] This crossover temperature is also substantially above the glass transition temperature. Wolynes and coworkers used the random first-order theory (RFOT) to suggest that two free energy barrier distributions merge (or bifurcate) at the crossover temperature.[10,37,57,58,62] The relaxation mechanism in the secondary (or β) relaxation occurs through strings. A relation between $T_B$ and $T_{cr}$, if indeed exists and if could be established, shall advance our understanding of the origin of anomalous dynamics significantly.

A quantitative understanding of the Johari-Goldstein bifurcation is partly plagued by the fact that both in computer simulation and in theoretical studies, one mostly employs models of spherical molecules,[6,39–41,71,72] although experimental systems involve molecules that are often highly non-spherical. Among the models studied, the Kob-Andersen (KA) model (a binary mixture of Lennard-Jones (LJ) spheres) has been widely employed in computer simulations of supercooled atomic liquids.[39–41] In this model, one introduces a certain size ratio and disparate interactions as additional complexity (compared to that of one-component systems) to prevent crystallization. As already mentioned above, one major problem remains, however. Most of the experimental studies, such as dielectric relaxation, nuclear magnetic resonance (NMR), and infrared (IR) spectra, primarily probe the orientational motion of molecules.[73] As aspherical molecules can rotate to facilitate molecular motion, this extra channel needs attention. Additionally, the aspherical molecules can be packed more densely than spherical molecules, as discussed in more detail below.[74,75] More recently, a study of glassy dynamics in a liquid of prolate-shaped colloidal particles has been reported.[76]

The density (or the packing fraction) of closely and randomly packed ellipsoids has been studied extensively.[74,75,77] Within certain aspect ratios not too different from unity, ellipsoids can pack better than spheres, giving higher density. The maximum packing fraction occurs at an aspect ratio of 0.6 for oblate spheroids and 1.80 for prolate spheroids.[75] A well-studied problem is the packing of M&M candies, which are oblates, and they pack in the disordered state better than spheres with a packing fraction close to 0.68.[75] It is well-known that the random closed pack (RCP) state of hard spheres can achieve a maximum packing fraction in the range of 0.62 to 0.64.[78] This packing fraction played an important role in our understanding of hard sphere glass transition or jamming transition. For hard prolate ellipsoidal, the estimate of the maximum packing fraction of random close packing ranges between 0.68 and 0.71.[75] This enhancement of packing fraction by 8-10% is expected to play a major role in the dynamics of ellipsoidal near the glass transition.

In fact, molecular shape and molecular geometry increasingly determine physical properties as the density of the liquid is increased. In particular, aspherical molecules can rotate and translate simultaneously to achieve a significant displacement. As discussed elsewhere, a prolate molecule exhibits faster diffusion along its long axis.[76,79,80] This translation-rotation coupling is absent in spherical molecules. Thus, the models employing spherical molecules lack certain fundamental aspects of the molecular dynamics of dense and supercooled liquids. These studies fail to address and reproduce many of the features observed near the glass transition. For example, in a classic study, Cicerone and Ediger observed that as the glass transition temperature was approached, the tagged non-spherical molecule translated further and further during its one rotational correlation time.[81,82] This was interpreted as arising from a decoupling of translational motion from viscosity due to dynamic heterogeneity. *While decoupling of translational diffusion from viscosity has been observed in models of spherical molecules,*[83] *the issue of the continued coupling of rotation to viscosity has not yet been satisfactorily resolved.*

The role of rotation in relaxation was demonstrated in a study of a system where an isolated ellipsoid was immersed in a sea of glass-forming liquid.[84] It was observed that rotation of the tagged ellipsoid could lead to the relaxation of the local stress tensor, thus lowering the local viscosity. Therefore, two opposite factors play out in a glassy liquid consisting of aspherical molecules: the ellipsoidal shape allows better packing so that glass transition can occur at higher density, whereas rotational motion can open up additional relaxation channels of stress relaxation and also transport. Thus, it could be fascinating to investigate the dynamics of molecules with orientational degrees of freedom across the supercooled regime.

In this work, we introduce a new class of glass-forming liquids that are binary mixtures of aspherical molecules (prolates and oblates). We present several new results. We find that the translational and rotational relaxation spectrum splits into two peaks at a temperature $T_B$, which is near, $T = 0.6$, while the glass transition temperature is 0.43. We identify this with the α-β bifurcation. Analysis of the individual peaks reveals fascinating insights. The slow peak shows greater heterogeneity. We attribute this to the presence of two different energy landscapes with varying roles of entropic and enthalpic stabilization. The product of diffusion constant $(D)$ and rotational relaxation time $(\tau_2)$ exhibits a rather sharp peak when plotted against temperature, at a given temperature $T_{cr}$ which is close to $T_B$, enabling us to establish a quantitative relation between the two temperatures. Furthermore, we coarse-grain the system into cubic boxes to study dynamic properties, revealing abrupt changes in the diffusion coefficient and rotational correlation time, which indicate first-order transitions between low and high-mobility domains.

The organization of the rest of the paper is as follows. In Section II, we discuss the model and technical details of the simulations. In Section III, we present a detailed discussion of the results obtained from our simulations. In this section, we first discuss the thermodynamic properties and relaxation dynamics of the model across the glass transition temperature. Subsequently, we discuss the bifurcation in the translational as well as rotational relaxation spectrum and its potential relationship with existing theoretical frameworks such as the Johari-Goldstein (JG) α-β bifurcation or that predicted by the Mode-Coupling Theory (MCT). Finally, we discuss the jump dynamic transitions in the mosaics. In Section IV, we summarize the present study and conclude with a few comments.

## II. MODEL AND SIMULATION DETAILS

We have performed extensive molecular dynamics (MD) simulations of the binary mixture comprising a total of 4000 particles with 3200 prolates (rod-shaped molecules) and 800 oblate ellipsoids (disc-shaped molecules) contained in a cubic box having periodic boundary conditions using Large-scale Atomic/Molecular Massively Parallel Simulator (LAMMPS) package.[85] Our decision to utilize a binary mixture of anisotropic molecules stemmed from our previous exploration of the widely studied Kob-Andersen (KA) spherical model to incorporate orientation into the model while retaining fundamental parameters for comparison. Binary mixtures provide a unique opportunity to examine how diverse particle characteristics, such as shape or size, influence overall system behaviour, enabling the exploration of complex scenarios beyond the single-particle behaviour seen in monodisperse systems.[86–89]

In this model system, the interactions between any two ellipsoids with arbitrary orientations are governed by a Gay-Berne (GB) potential.[90,91] In the Gay-Berne pair potential, each ellipsoid of revolution $i$ is represented by the position $\mathbf{r}_i$ of its centre of mass and a unit vector $\mathbf{e}_i$ along the principal symmetry axis, as shown in **Figure 1(a)** for a pair of prolates. The Gay-Berne potential for the interaction between two ellipsoids of revolution $i$ and $j$ is given by the following expression,[90,91]

$$U_{ij}^{GB}\left(\mathbf{r}_{ij}, \mathbf{e}_i, \mathbf{e}_j\right) = 4\varepsilon_{ij}\left(\hat{\mathbf{r}}_{ij}, \mathbf{e}_i, \mathbf{e}_j\right)\left(\rho_{ij}^{-12} - \rho_{ij}^{-6}\right) \tag{1}$$

where,

$$\rho_{ij} = \frac{r_{ij} - \sigma\left(\hat{\mathbf{r}}_{ij}, \mathbf{e}_i, \mathbf{e}_j\right) + \sigma_{GB}}{\sigma_{GB}}. \tag{2}$$

Here, $\sigma_{GB}$ represents the cross-sectional diameter along the breadth, $r_{ij}$ is the distance between the two centres of mass, and $\hat{\mathbf{r}}_{ij} = \mathbf{r}_{ij}/r_{ij}$ is a unit vector along the intermolecular separation vector $\mathbf{r}_{ij}$. The molecular shape parameter $\sigma(\hat{\mathbf{r}}_{ij}, \mathbf{e}_i, \mathbf{e}_j)$ is given by the expression,

$$\sigma(\hat{\mathbf{r}}_{ij}, \mathbf{e}_i, \mathbf{e}_j) = \sigma_{GB}\left[1 - \frac{\chi}{2}\left\{\frac{(\mathbf{e}_i \cdot \hat{\mathbf{r}}_{ij} + \mathbf{e}_j \cdot \hat{\mathbf{r}}_{ij})^2}{1 + \chi(\mathbf{e}_i \cdot \mathbf{e}_j)} - \frac{(\mathbf{e}_i \cdot \hat{\mathbf{r}}_{ij} - \mathbf{e}_j \cdot \hat{\mathbf{r}}_{ij})^2}{1 - \chi(\mathbf{e}_i \cdot \mathbf{e}_j)}\right\}\right]^{-1/2} \quad (3)$$

where, $\chi = \dfrac{(\kappa^2 - 1)}{(\kappa^2 + 1)}$. Here $\kappa$ denotes the aspect ratio of the ellipsoid of revolution and is given by $\kappa = \sigma_{ee}/\sigma_{ss}$. $\sigma_{ee}$ is the molecular length along the principal symmetry axis and $\sigma_{ss} = \sigma_{GB}$. The energy parameter $\varepsilon(\hat{\mathbf{r}}_{ij}, \mathbf{e}_i, \mathbf{e}_j)$ is given by the expression,

$$\varepsilon_{ij}(\hat{\mathbf{r}}_{ij}, \mathbf{e}_i, \mathbf{e}_j) = \varepsilon_{GB}\left[\varepsilon_1(\mathbf{e}_i, \mathbf{e}_j)\right]^\nu \left[\varepsilon_2(\hat{\mathbf{r}}_{ij}, \mathbf{e}_i, \mathbf{e}_j)\right]^\mu \quad (4)$$

where, $\mu$ and $\nu$ are two exponents which are adjustable, and

$$\varepsilon_1(\mathbf{e}_i, \mathbf{e}_j) = \left[1 - \chi^2 (\mathbf{e}_i \cdot \mathbf{e}_j)^2\right]^{-1/2} \quad (5)$$

and,

$$\varepsilon_2(\hat{\mathbf{r}}_{ij}, \mathbf{e}_i, \mathbf{e}_j) = 1 - \frac{\chi'}{2}\left[\frac{(\mathbf{e}_i \cdot \hat{\mathbf{r}}_{ij} + \mathbf{e}_j \cdot \hat{\mathbf{r}}_{ij})^2}{1 + \chi'(\mathbf{e}_i \cdot \mathbf{e}_j)} + \frac{(\mathbf{e}_i \cdot \hat{\mathbf{r}}_{ij} - \mathbf{e}_j \cdot \hat{\mathbf{r}}_{ij})^2}{1 - \chi'(\mathbf{e}_i \cdot \mathbf{e}_j)}\right]. \quad (6)$$

Here, $\chi' = (\kappa'^{1/\mu} - 1)/(\kappa'^{1/\mu} + 1)$ with $\kappa' = \varepsilon_{ss}/\varepsilon_{ee}$. $\varepsilon_{ss}$ (or $\varepsilon_{GB}$) represents the depth of the minimum of the potential for a pair of ellipsoids when they are aligned side-by-side, and $\varepsilon_{ee}$ is the corresponding depth for end-to-end alignment.

The Gay-Berne potential defines a family of potential, each member of which is characterized by a set of four parameters $(\kappa, \kappa', \mu \text{ and } \nu)$. $\kappa$ being the aspect ratio, provides a measure of the shape anisotropy while $\kappa'$ provides a measure of the anisotropy of the well depth, which can also be controlled by the two parameters $\mu$ and $\nu$. In our study, the prolate is characterized by these parameters as $(1.2, 3, 2\ and\ 1)$, while the oblate is characterized by $(0.8, 3, 2\ and\ 1)$.

Furthermore, correspond to the aspect ratio of prolates and oblates (i.e., 1.2 and 0.8, respectively), $\sigma_{ee}$ and $\sigma_{ss}$ are chosen such that the volume of a prolate is equal to the volume of an oblate. A schematic representation of the functional form of the Gay-Berne potentials characterized by a set of parameters (1.2, 3, 2, 1) is shown in **Figure 1(b).**

All the quantities presented in this work are in reduced units and are defined in terms of the Gay-Berne potential parameters, $\sigma_{GB}$ and $\varepsilon_{GB}$: for example, length in units of $\sigma_{GB}$, temperature in units of $\varepsilon_{GB}/k_B$, $k_B$ being the Boltzmann constant, and time in units of $\left(m\sigma_{GB}^2/\varepsilon_{GB}\right)^{1/2}$, m being the mass of ellipsoids. We set the masses of both prolates and oblates equal to unity. The energy parameters of the interaction potential for our study have been chosen as: $\varepsilon_{PP} = 1.0, \varepsilon_{OO} = 0.5$ and $\varepsilon_{PO} = 1.5.$ It is to be noted that the choice of the energy parameters is inspired by those of the Kob-Andersen binary mixture.[39–41] For argon, these units correspond to $\sigma = 3.405 \times 10^{-10}$ m, $\varepsilon/k_B = 119.8$ K and $m = 0.03994$ kg/mol.

The system has been melted from a fcc configuration at a high temperature and low density, and then gradually compressed to a volume corresponding to the desired temperature and pressure. We have investigated the system for a series of reduced temperatures ranging from $T$ = 1.0 to 0.30 upon gradual cooling. The change of temperature has been effected by coupling the system to the Nosé–Hoover thermostat[92] at the required temperature for a time period of $t_{eq}$ that has been chosen to be larger than the relaxation time of the system at that temperature. The pressure of the system is kept constant, in all cases, at $P$ = 30 using the Parrinello–Rahman barostat.[93] Our rationale for selecting high-pressure conditions in our simulations is to address challenges arising from the pressure dependence of the glass transition temperature. Operating at lower pressures could potentially lead to the glass transition occurring at temperatures that are exceedingly difficult to reach within a reasonable simulation time frame.[94] Therefore, opting for high-pressure conditions ensures the glass transition occurs at more feasible temperatures, striking a balance between computational efficiency and accurate representation of dynamics and structural properties in the glassy state.

First of all, we have performed energy minimization using the steepest descent algorithm[95] followed by initial runs in the isothermal-isobaric (NPT) ensemble for $10^8$ steps. The production runs have been performed in the isothermal-isobaric (NPT) ensemble for $5 \times 10^8$ steps. The equations of motion have been integrated using the velocity-Verlet algorithm with integration time steps of dt = 0.001 for higher temperatures ($T \geq 0.6$) and dt = 0.002 for lower temperatures ($T < 0.6$). The potential has been cut off and shifted at a distance of $4.0\sigma_{GB}$. To improve the statistics, we have carried out three independent runs corresponding to each temperature studied. We report here the results averaged over these runs.

Besides the usual molecular dynamics (MD) simulations, we have also performed simulations in which the system is subjected to cooling (from $T = 1.0$ to $0.25$) and subsequent reheating at a constant rate to get insight into the thermodynamic properties of the system across glass transition temperature. The system has been subjected to cooling-reheating cycles for three different constant cooling/heating rates of $2.5 \times 10^{-8}$, $1.0 \times 10^{-7}$ and $4.0 \times 10^{-7}$.

## III. RESULTS AND DISCUSSION

### A. Thermodynamic properties

A glass transition is usually marked by a sharp, although continuous, change in thermodynamic properties such as volume (or density) and enthalpy, as the temperature is varied across the glass transition region.[6,7,24] After the glass transition, the thermodynamic quantities attain a value comparable to that of a crystalline solid. In order to explore the thermodynamic aspects of the system under study, first of all, we calculate the specific volume (defined as the inverse of number density) at different temperatures using molecular dynamics (MD) simulation in the isothermal-isobaric (NPT) ensemble. The variation of specific volume $(V_{sp})$ as a function of temperature is shown in the Supplementary Material (Figure S1). The temporal evolution of specific volume changes its slope, indicating the system undergoes glass transition. The intersection of the curves (when extrapolated) on both sides of the transformation range predicts the glass transition temperature $T_g = 0.43$. In the work done by Kimura and Yonezawa, a change in the slope of specific volume with respect to temperature was observed near the glass transition of spherical molecules.[6,96] In **Figure 1(c),** we plot the thermal expansion coefficient, defined as $\alpha = \frac{1}{V_{sp}} \left( \frac{\partial V_{sp}}{\partial T} \right)_P$, as a function of temperature. The thermal expansion coefficient $(\alpha)$ shows a sharp change near the glass transition temperature $T_g = 0.43$.

Besides the specific volume, another important quantity that is widely used to study the glass transition is isobaric heat capacity. In calorimetric experiments, a rapid decrease in isobaric heat capacity is usually observed when a glass-forming liquid is subjected to isobaric cooling. The isobaric heat capacity is defined as[97]

$$C_p = \left(\frac{\partial \langle H \rangle}{\partial T}\right)_P = \frac{1}{k_B T^2} \sigma_H^2 \tag{7}$$

where, $\langle H \rangle$ is the average enthalpy of the system at temperature $T$, and $\sigma_H^2 = \left(\langle H^2 \rangle - \langle H \rangle^2\right)$.

In **Figure 1(d)**, we show the variation of the isobaric heat capacity as a function of temperature. Similar to the thermal expansion coefficient, we also observe a sharp change in the isobaric heat capacity, which is again a signature of glass transition.

In addition to the usual MD simulations, we have calculated the specific volume, thermal expansion coefficient, and isobaric heat capacity via the isobaric cooling at $P = 30$ (from $T = 1.0$ to $0.25$) and subsequent reheating of the system at constant rates. In **Figures 1(c) and 1(d)**, we also show the variation of the thermal expansion coefficient and isobaric heat capacity upon cooling and reheating for three different constant cooling/heating rates $\left(2.5 \times 10^{-8}, 1.0 \times 10^{-7} \text{ and } 4.0 \times 10^{-7}\right)$. We find that our model system shows hysteresis in the isobaric heat capacity and the thermal expansion coefficient during cooling and subsequent reheating, as experimentally observed in the glass-forming liquids.[98,99] Further, on varying the cooling/heating rate, we find that the dependence of $T_g$ on the cooling/heating rate is weak, which is consistent with the earlier predictions.[100,101]

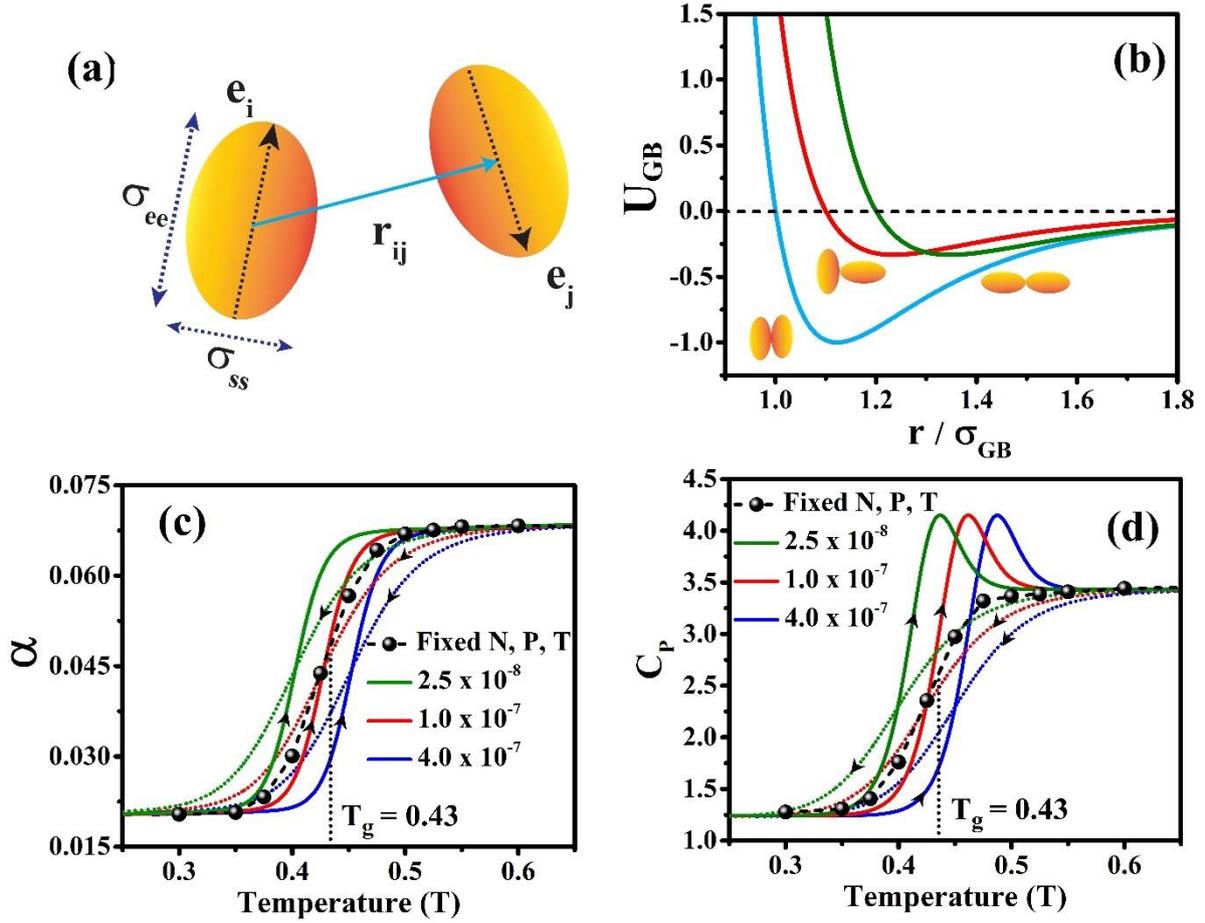

**Figure 1.** (a) Schematic diagram of a pair of ellipsoids defined by Gay-Berne (GB) potential parameters, and (b) functional form of the Gay-Berne pair potential characterized by a set of parameters (1.2, 3, 2, 1). The ratio of the energy depth of the side-by-side configuration (corresponding to the deepest energy depth, blue line) to that of the end-to-end configuration (corresponding to the shallowest energy depth, green line) shows the value of $\kappa' = 3.0$. The temperature dependence of (c) the thermal expansion coefficient and (d) isobaric heat capacity. The results are obtained by MD simulations in the NPT ensemble (black spheres with dashed line), isobaric cooling (dotted lines), and isobaric reheating (solid lines) of the system for three different constant cooling/heating rates. A sharp change in the thermal expansion coefficient and isobaric heat capacity shows a signature of glass transition, with the glass transition temperature at $T_g = 0.43$ (from MD simulation in the NPT ensemble, as indicated by a black dotted line). Further, these properties show hysteresis during cooling and subsequent reheating.

One often uses the Prigogine-Defay ratio (PDR) to characterize the intensity of a glass transition. It is a combination of the change in response functions involving specific heat, isothermal compressibility, and thermal expansion coefficient during the glass transition. The Prigogine-Defay ratio (PDR), usually denoted by $\Pi$ and given by[102,103]

$$\Pi = \frac{1}{V_{sp} T} \left\{ \frac{\Delta C_P \Delta \kappa}{(\Delta \alpha)^2} \right\}\bigg|_{T=T_g} \quad (8)$$

where, $\Delta C_P$ is the change in isobaric heat capacity, $\Delta \kappa$ is the jump of isothermal compressibility, and $\Delta \alpha$ is the change in thermal expansion coefficient near the glass transition temperature $T = T_g$. The variation of isothermal compressibility (obtained from MD simulations in NPT ensemble) as a function of temperature across the glass transition region is given in the Supplementary Material (Figure S2).

An important aspect of PDR is that all the terms involved are the response functions themselves. For the present system, we can calculate the change in response functions and, thus, the value of the PDR. For molecular liquids, the value of PDR is typically in the range of 1-10, with glycerol having a value of 9.4 and propanol 1.9.[103] We have obtained a PDR value of 2.7 for our model system at the glass transition temperature (from MD simulations in NPT ensemble), which appears to be in the correct range.

In the subsequent sections, we focus on the dynamics of the system.

**B. Relaxation dynamics**

In this section, we focus on the translational and rotational dynamics of the glass-forming model system of anisotropic molecules. It is to be noted that, at the single particle level, both the components of the binary mixture are dealt with individually.

First, we look into the translational dynamics of the glass-forming model systems by evaluating the self-part of the intermediate scattering function. The self-intermediate scattering function is given by the expression[104]

$$F_s(k,t) = \left\langle \exp\left[i\mathbf{k}.\left(\mathbf{r}_j(t) - \mathbf{r}_j(0)\right)\right] \right\rangle \quad (9)$$

where, $\mathbf{r}_j(t)$ is the position of the center of mass of $j^{th}$ particle at time $t$.

The time evolution of $F_s(k,t)$ at $k \sim k_{max}$, ($k_{max}$ is the position of the dominant peak of the static structure factor $S(k)$) for prolates and oblates at different temperatures is shown in **Figures 2(a) and 2(b),** respectively. At high temperatures, the long time decay of $F_s(k,t)$ is essentially exponential in nature; however, as the temperature is lowered, the decay of $F_s(k,t)$ distinctly involves the separation of timescales with an intervening plateau. We shall come to this point later in the text.

We can get an estimation of the time scale of the structural relaxation from the decay of $F_s(k,t)$. We determine the α-relaxation time $\tau_\alpha$ as, $F_s(k,\tau_\alpha) = e^{-1}$. We fit the α- relaxation time of $F_s(k,t)$ by using Vogel-Fulcher-Tammann (VFT) equation, which is as follows,[105]

$$\tau_\alpha(T) = \tau_0 \exp\left[\frac{A}{T-T_g}\right] \qquad (10)$$

Here, $\tau_0$ is a constant, the parameter $A$ is related to the fragility, $D = A/T_g$ and $T_g$ is the temperature of the "ideal" glass transition, obtained by Vogel-Fulcher-Tammann (VFT) fit. **Figure 3(a)** depicts the variation of α- relaxation time $(\tau_\alpha)$ as a function of $T_g/T$ for prolates and oblates. It also shows the VFT fit of $\tau_\alpha$ for prolates and oblates. The VFT fit predicts the glass transition temperature, $T_g = 0.430$ for prolates and $T_g = 0.434$ for oblates. The values of $T_g$ obtained by VFT fit are in good agreement with that obtained by thermodynamic properties.

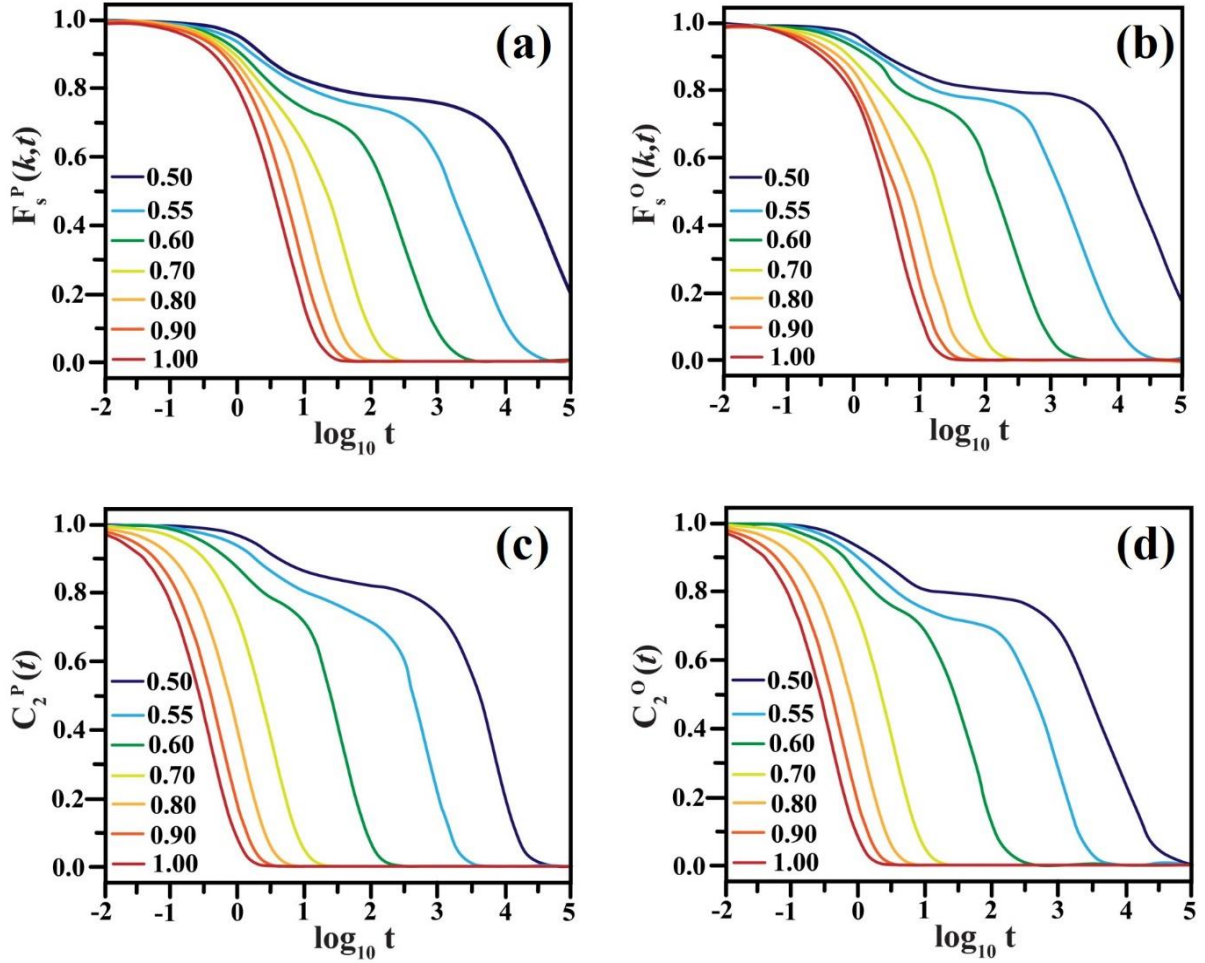

**Figure 2. Temporal evolution of the self-part of the intermediate scattering function $F_s(k \sim k_{max}, t)$ of (a) prolate and (b) oblate ellipsoids of revolution for different temperatures investigated. At low temperatures, the decay distinctly involves the separation of time scales with an intervening plateau. The time evolution of the single-particle 2$^{nd}$ rank orientational relaxation function of (c) prolates and (d) oblates at different temperatures investigated.**

The divergence of α-relaxation time $(\tau_\alpha)$ near the glass transition is also predicted as a power law by the ideal mode-coupling theory (MCT) for a particular range of temperatures. The power-law behaviour can be observed by fitting the α-relaxation time $(\tau_\alpha)$ using the following equation,[40,41]

$$\tau_\alpha(T) = C(T - T_c)^{-\gamma} \qquad (11)$$

Here, C is a prefactor and $\gamma$ is a constant for a system; $T_c$ is the critical temperature predicted by ideal MCT at which the system undergoes a transition from an ergodic state to a non-ergodic state.

**Figure 3(b)** depicts the power-law behaviour of the α-relaxation time $(\tau_\alpha)$. The power law fit predicts the critical temperature, $T_c = 0.546$ and the exponent $\gamma = 1.96$ for prolates, while for oblates, it provides the value of $T_c = 0.548$ and $\gamma = 1.90$. According to MCT, the exponent $\gamma$ and the critical temperature $T_c$ should be independent of particle type. The deviation observed here is comparable to that observed before for the Kob-Andersen binary mixture and should not be considered as a severe contradiction to the prediction of MCT.[40] For both prolates and oblates, we observe that $T_g < T_c$, which is in agreement with the previous studies.[40,41]

As discussed earlier, the decay of $F_s(k,t)$ at lower temperatures involves both fast and slow timescales with an intervening plateau. Both the approach to the plateau and the decay from the plateau are given by power laws, with exponents *a* and *b*, as described below. The short-time decay to the plateau is governed by the power law,

$$F_s(k,t) \sim f + At^{-a} \tag{12}$$

whereas, the decay from the plateau region is governed by another power law (known as the von Schweidler law),

$$F_s(k,t) \sim f - Bt^b \tag{13}$$

where, *f* is the plateau height, *A* and *B* are constants. Near the MCT critical temperature $T_c$, the exponents '*a*' and '*b*' are related as,

$$\frac{[\Gamma(1-a)]^2}{\Gamma(1-2a)} = \frac{[\Gamma(1+b)]^2}{\Gamma(1+2a)} \tag{14}$$

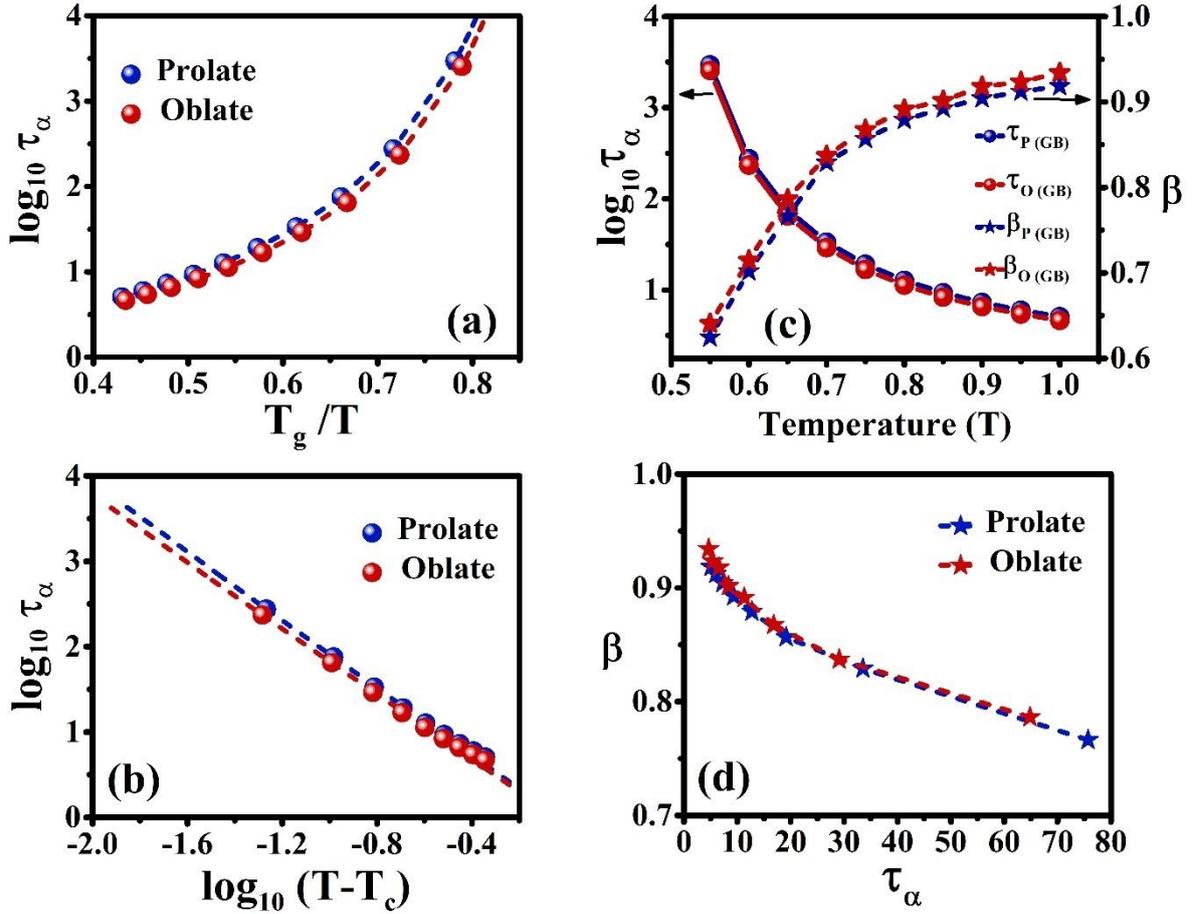

**Figure 3.** (a) α-relaxation time ($\tau_\alpha$) of prolates and oblates as a function of $T_g/T$, where $T_g$ is the glass transition temperature predicted by the Vogel-Fulcher-Tammann (VFT) fit. The blue and red dashed lines correspond to the VFT fits of the α- relaxation times $\tau_\alpha$ of prolates and oblates, respectively. Such fits yield $T_g = 0.430, \tau_0 = 0.934, A = 0.967$ (for prolates) and $T_g = 0.434, \tau_0 = 0.912, A = 0.921$ (for oblates). (b) α- relaxation time ($\tau_\alpha$) as a function of temperature difference $(T - T_c)$, where $T_c$ is the mode coupling critical (or divergence) temperature. The blue and red dashed lines represent the power law fits of $\tau_\alpha$ with $T_c = 0.546$, $\gamma = 1.96$ for prolates and $T_c = 0.548$, $\gamma = 1.90$ for oblates, respectively. (c) The variation of the relaxation time $\tau_\alpha$ and the exponent $\beta$ (obtained from Kohlrausch-Williams-Watts stretched exponential fit of $F_s(k,t)$) as a function of temperature for prolates and oblates. *The most striking result is the signature of the strong anti-correlation between $\tau_\alpha$ and $\beta$ values.* (d) The variation of the exponent $\beta$ with the relaxation time $\tau_\alpha$. The lines are drawn as a guide to the eye.

where, $\Gamma$ denotes the Gamma function. Further, the power law exponents $'a'$ and $'b'$ are related to the constant $\gamma$ (described in Eq. 11) by,

$$\gamma = \frac{1}{2a} + \frac{1}{2b}. \tag{15}$$

In **Table 1**, we show the value of exponents $'a'$ and $'b'$ obtained by fitting of $F_s(k,t)$ at $T = 0.55$ (near the critical temperature $T_c$ predicted by MCT). The fittings are shown in the Supplementary Material (Figure S3). Here, we note that gamma is indeed quite close to the prediction of MCT, as has been observed in other simulations.

**Table1: The value of exponents $'a'$ and $'b'$ obtained by fitting of $F_s(k,t)$ at T = 0.55 (near the critical temperature predicted by MCT). We also provide the value of the exponent $\gamma$ obtained by eqn. (15).**

|  | $a$ | $b$ | $\gamma$ |
|---|---|---|---|
| **Prolate** | 0.37 | 0.82 | 1.96 |
| **Oblate** | 0.39 | 0.81 | 1.89 |

Finally, at sufficiently long times, the fall from the plateau occurs, which is known as $\alpha$-relaxation. The final relaxation, i.e., $\alpha$–relaxation regime, is predicted to follow the Kohlrausch-Williams-Watts (KWW) stretched exponential form, which is as follows:

$$F_S(k \sim k_{\max}, t) \sim \exp\left[-\left(t/\tau_\alpha(T)\right)^{\beta(T)}\right] \tag{16}$$

**Figure 3(c)** depicts the variation of the exponent $\beta(T)$ and relaxation time $\tau_\alpha$ obtained from the stretched exponential fit of $F_s(k \sim k_{\max}, t)$, as a function of temperature. For the system under study, we observe an anti-correlation between the exponent $\beta(T)$ and relaxation time $\tau_\alpha$. In **Figure 3(c),** the relaxation time is seen to grow rapidly as the glass transition temperature is approached from above. By fitting the divergence of the relaxation time, we find a value of $T_g = 0.43$. The exponent $\beta$, on the other hand, decreases towards 0.6 in the same range of temperature. These two variations are in excellent agreement with known results.[40,41]

In **Figure 3(d)**, we present the variation of the exponent $\beta$ against the relaxation time $\tau_\alpha$; both have been obtained by fitting our simulation data to the Kohlrausch-Williams-Watts (KWW) stretched exponential form, Eq. (16). It is seen that both prolates, and oblates follow the same dependence. The decrease in the value of the exponent implies that the relaxation becomes increasingly heterogeneous. The increase in the relaxation time indicates the growth of slow glassy regions in the system. These features are well-known and relatively easy to understand. The main point here is that our new model of glass-forming liquids with molecular shapes reproduces most of the known experimental results.

Besides the self-intermediate scattering function, another important property of interest is the self-diffusion coefficient, which can capture the slowdown of translational dynamics as the temperature falls. The translational self-diffusion coefficient of component $A$ (say prolate or oblate) has been obtained from the mean squared displacement (MSD) by the use of the Einstein relation,

$$D_A = \lim_{t \to \infty} \frac{\langle \Delta \mathbf{r}_A^2(t) \rangle}{6t} \quad (17)$$

where,

$$\langle \Delta \mathbf{r}_A^2(t) \rangle = \frac{1}{N_A} \sum_{i=1}^{N_A} \langle |\mathbf{r}_i(t) - \mathbf{r}_i(0)|^2 \rangle \quad (18)$$

$\mathbf{r}_i(t)$ being the position of the centre of mass of an $i^{th}$ particle at time $t$. Here, $A$ can be prolate or oblate; accordingly, the sum goes either over the prolates or oblates.

The temporal evolution of mean square displacements (MSD) of prolates and oblates for different temperatures investigated is shown in the Supplementary Material (Figure S4). The diffusion constants are obtained by the linear fitting of mean square displacement versus time curves (by avoiding the initial ballistic region). In **Figure 4(a),** we plot the variation of the calculated self-diffusion coefficient of both the prolate and the oblate particles as we lower the temperature towards the glass transition temperature. *We observed almost four orders of magnitude decrease in the value of the self-diffusion coefficient D.* Further, the prolates and the oblates both follow a similar temperature dependence. In **Figure 4(a)**, we also show the VFT fit of the self-diffusion coefficient (*D*) for prolates and oblates (shown by dashed lines), using the relation, $D(T) = D_0 \exp\left[-\frac{A}{T-T_g}\right]$. The VFT fit predicts the glass transition temperature, $T_g = 0.431$ for prolates and $T_g = 0.433$ for oblates. The values of $T_g$ obtained by VFT fit are in good agreement with those obtained by other properties.

Similar to the α-relaxation time $(\tau_\alpha)$, the divergence (or disappearance) of the translational diffusion coefficient *D,* near the glass transition, is also predicted as a power law by the ideal mode-coupling theory (MCT) for a particular range of temperature. The power-law behaviour can be observed by fitting the translational diffusion coefficient *D* using the following equation,

$$D(T) = C\left(T - T_c\right)^\gamma \quad (19)$$

In **Figure 4(b),** we show the power-law behaviour of the translational diffusion coefficient *D* of prolates and oblates. For each of the components, the translational diffusion is found to follow a power law dependence as predicted by the ideal MCT. The power law fit predicts the critical temperature, $T_c = 0.545$ and the exponent $\gamma = 1.92$ for prolates, while for oblates, it provides the value of $T_c = 0.547$ and $\gamma = 1.84$.

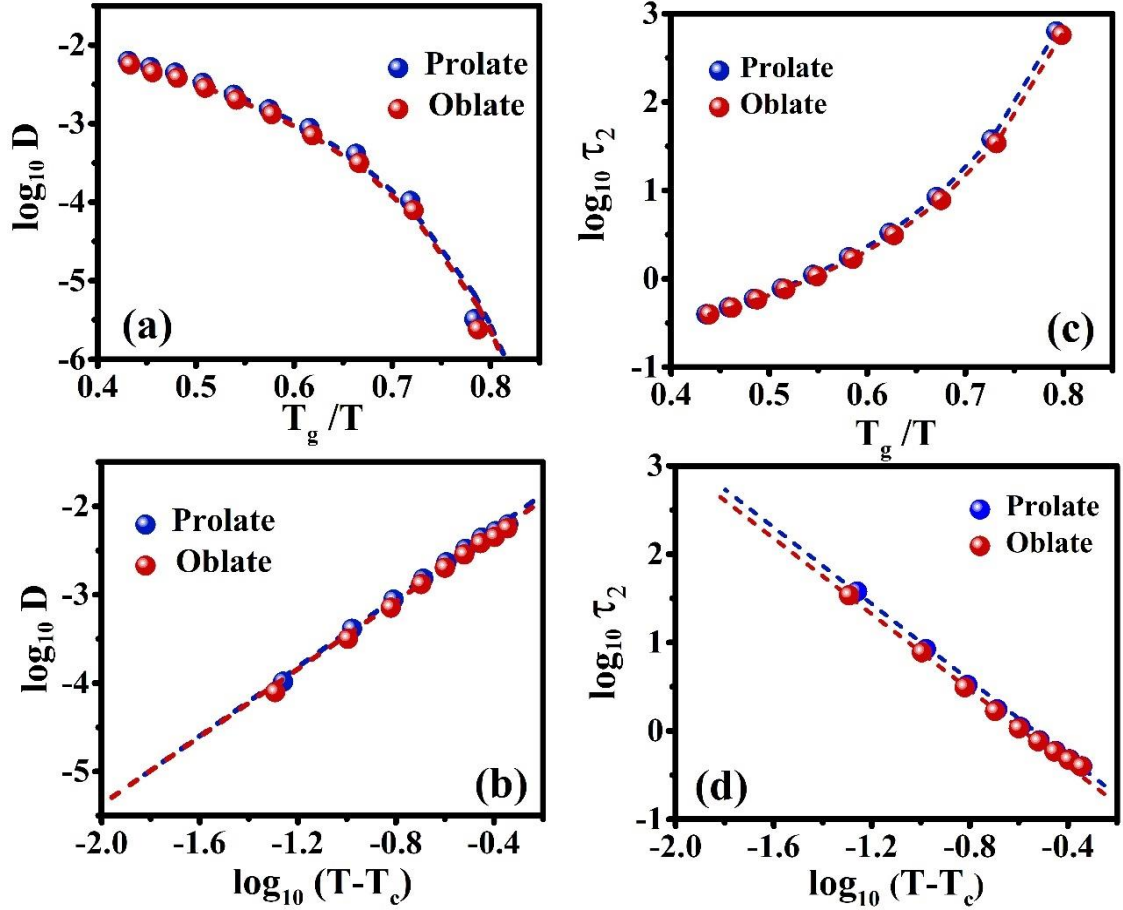

**Figure 4.** (a) Translational diffusion coefficient, $D$, of prolates and oblates as a function of $T_g/T$. The blue and red dashed lines correspond to the VFT fits of the translational diffusion coefficient of prolates and oblates, respectively. Such fits yield $T_g = 0.431$, $D_0 = 0.038$, $A = 1.027$ for prolates and $T_g = 0.433$, $D_0 = 0.035$, $A = 1.041$ for oblates. (b) Translational diffusion coefficient as a function of temperature difference $(T - T_c)$, where $T_c$ is the mode coupling critical (or divergence) temperature. The blue and red dashed lines represent the power law fits of the translational diffusion coefficient with $T_c = 0.545$, $\gamma = 1.92$ for prolates and $T_c = 0.547$, $\gamma = 1.84$ for oblates, respectively. (c) The variation of the second-rank orientational relaxation time $(\tau_2)$ of prolates and oblates with $T_g/T$, where $T_g$ is the glass transition temperature predicted by the Vogel-Fulcher-Tammann (VFT) fit. The blue and red dashed lines correspond to the VFT fits of the translational diffusion coefficient of prolates and oblates, respectively. Such fits yield $T_g = 0.433$, $\tau_{2,0} = 0.062$, $A = 1.053$ for prolates and $T_g = 0.435$, $\tau_{2,0} = 0.056$, $A = 1.092$ for oblates. (d) Second rank relaxation time as a function of temperature difference $(T - T_c)$, where $T_c$ is the mode coupling critical (or divergence) temperature. The blue and red dashed lines represent the power law fits $\tau_2$ of the with $T_c = 0.545$, $\gamma = 1.94$ for prolates and $T_c = 0.548$, $\gamma = 1.89$ for oblates, respectively.

In order to get an insight into the rotational dynamics of the particles, we calculate the second-rank orientational relaxation time, $(\tau_2)$ which is obtained by the integration of the second-rank orientational time correlation function. The second-rank orientational time correlation function, $C_2(t)$, is defined as[70]

$$C_2(t) = \frac{\left\langle \sum_{i=1}^{N_A} P_2\left(\mathbf{e}_i(t) \cdot \mathbf{e}_i(0)\right) \right\rangle}{\left\langle \sum_{i=1}^{N_A} P_2\left(\mathbf{e}_i(0) \cdot \mathbf{e}_i(0)\right) \right\rangle} \quad (20)$$

where, $\mathbf{e}_i$ is a unit vector along the principal symmetry axis of the ellipsoid. In this context, $A$ can be either prolate or oblate; thus, the sum is performed over the number of prolates or oblates. $P_2(x)$ is the second-order Legendre polynomial, expressed as

$$P_2(x) = \frac{1}{2}\left(3x^2 - 1\right) \quad (21)$$

The second-rank orientational time correlation function, $C_2(t)$ is directly accessible to experiments (for example, NMR and ESR provide information on $C_2(t)$).[73] We show the time evolution of the second-rank orientational time correlation function of prolates and oblates in **Figures 2(c) and 2(d).**

The second-rank orientational relaxation time is given by the expression,[70]

$$\tau_2 = \int_0^\infty C_2(t)dt \quad (22)$$

The variation of the second-rank orientational relaxation time $(\tau_2)$ of prolates and oblates as a function of $T_g/T$ is shown in **Figure 4(c)**. As the temperature is lowered, the decay of the orientational time correlation function becomes slower. We also fit the second-rank orientational relaxation time $(\tau_2)$ of prolates and oblates using the Vogel-Fulcher-Tammann (VFT) equation, $\tau_2(T) = \tau_{2,0} \exp\left[\dfrac{A}{T-T_g}\right]$. The VFT fit predicts the glass transition temperature, $T_g = 0.431$ for prolates and $T_g = 0.433$ for oblates. Similar to the α-relaxation time and self-diffusion coefficient, the power-law behaviour of the second-rank orientational relaxation time $(\tau_2)$ can be observed (**Figure 4(d)**) by fitting to the equation, $\tau_2(T) = C(T-T_c)^{-\gamma}$. Such fit yields $T_c = 0.545$, $\gamma = 1.92$ for prolates and $T_c = 0.548$, $\gamma = 1.89$ for oblates, respectively.

## C. α–β bifurcation

In order to get an insight into the heterogeneous dynamics of the system near the glass transition, we have calculated the distribution (or histogram) of the short-to-intermediate time diffusion constant of prolates and oblates. The short-to-intermediate time diffusion coefficient of each particle is obtained from the linear fit of each MSD curve (from $t = 10$ to $50$). In **Figures 5(a) to 5(f)**, we show the distribution of the short-to-intermediate time diffusion coefficient, $P(D,T)$, of prolates at six different temperatures ($T$ = 1.0, 0.8, 0.7, 0.6, 0.55, and 0.5). At high temperatures (T = 1.0 or so), the distribution curve exhibits a single peak. However, with decreasing temperature (from T = 0.8 to 0.7), the peak progressively broadens, eventually giving rise to a bimodal distribution at lower temperatures (at T = 0.6 and 0.55). The bimodal distribution indicates the presence of a mosaic-like structure in the deeply supercooled liquid phase. On further lowering the temperature, a single peak corresponding to a very low diffusion coefficient value is observed, indicating that the dynamics of the particle become sluggish. We obtained similar results when the distributions of the short-to-intermediate time diffusion coefficient of oblates were plotted, as shown in the Supplementary Material (Figure S5).

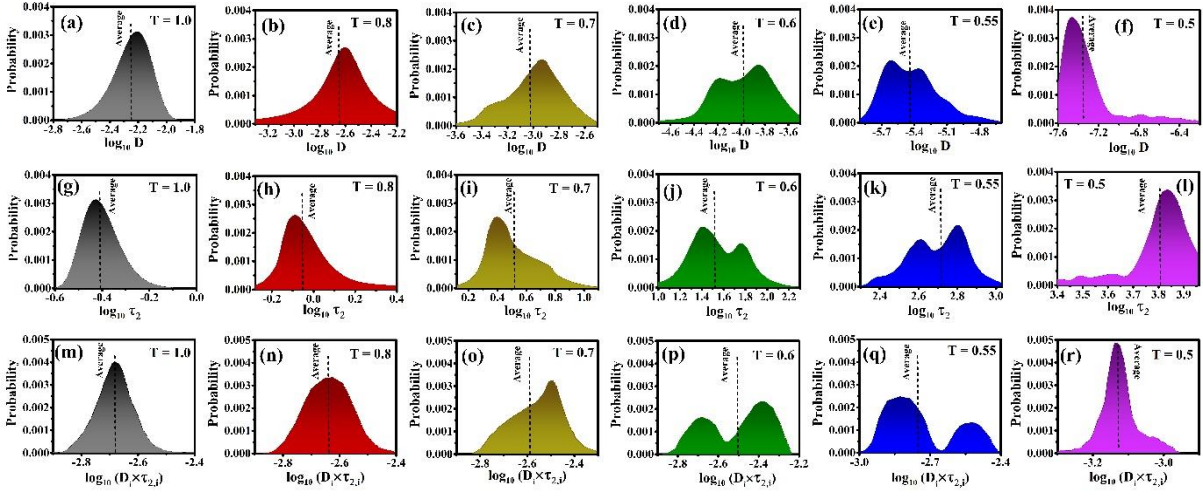

**Figure 5. The distribution of the short-to-intermediate time diffusion coefficient of prolates at (a) T = 1.0, (b) T=0.8, (c) T=0.7, (d) T=0.6, (e) T=0.55, and (f) T=0.5. The distribution of the intermediate time orientational relaxation times of prolates at (g) T=1.0, (h) T=0.8, (i) T=0.7, (j) T=0.6, (k) T=0.55 and (l) T=0.5. The distribution of $D_i \times \tau_{2,i}$ for each prolate particle at (m) T = 1.0, (n) T=0.8, (o) T=0.7, (p) T=0.6, (q) T=0.55, and (r) T=0.5. In all the cases, at a higher temperature, there is only one peak; however, as the temperature is lowered, a bimodal distribution is observed. It indicates the presence of a mosaic-like structure in the deeply supercooled liquid phase. As the temperature is further lowered, the dynamics of the particle become sluggish.**

Similar to the distribution of the short-to-intermediate time diffusion coefficient $P(D,T)$, the distribution of the intermediate-time orientational relaxation time $P(\tau_2,T)$ can also provide insight into the heterogeneous dynamics of the system. For this purpose, we define the intermediate-time orientational relaxation time such that it corresponds to the time at which the second-rank orientational time correlation function of particles becomes $e^{-1}$ i,e. $C_2(\tau_2) = e^{-1}$. In **Figures 5(g) to 5(l),** we show the distribution of the intermediate-time orientational relaxation time $P(\tau_2,T)$ of prolates at six different temperatures. It shows similar temperature dependence as seen in the distribution of short-to-intermediate time diffusion coefficient. Additionally, we have determined $P(D\tau_2,T)$ of $D_i \times \tau_{2,i}$ for each particle, as shown in **Figures 5(m) to 5(r).** Interestingly, similar to $P(D,T)$ and $P(\tau_2,T)$, it also exhibits a bifurcation as the temperature decreases.

What is depicted in **Figure 5** is a remarkable appearance of a bifurcation in the relaxation spectrum. While this is akin to the bifurcation observed and analyzed by Johari and Goldstein, the present bifurcations take place in the ps to ns time scales. This agrees rather well with the recent report of such bifurcation by Cicerone et al.[38] In the Johari-Goldstein α−β bifurcation phenomenon, the primary branch rapidly disappears in a markedly non-Arrhenius fashion when the glass transition is approached from above, leaving the secondary branch that exhibits Arrhenius-like temperature dependence.[31,32]

The appearance of the bifurcation seems to suggest the existence of two different basins in the energy landscape. We interpret them as a liquid-like basin (giving rise to the fast mode) and a solid-like basin (as the slow mode). In order to get insight into the dynamic heterogeneity in the two basins, we calculate the non-Gaussian parameter and the four-point dynamical susceptibility corresponding to the particles that constitute the slow and fast peaks of spectra at T= 0.60 and T=0.55.

The translational non-Gaussian parameter is defined as,[106]

$$\alpha_2(t) = \frac{3\langle \Delta \mathbf{r}(t)^4 \rangle}{5\langle \Delta \mathbf{r}(t)^2 \rangle^2} - 1 \tag{23}$$

If $\mathbf{r}(t)$ is a Gaussian process, $\alpha_2(t)$ vanishes for all time.

The four-point dynamic susceptibility is defined as follows,[107]

$$\chi_4(t) = \frac{1}{N}\left[\langle Q^2(t) \rangle - \langle Q(t) \rangle^2\right] \tag{24}$$

where,

$$Q(t) \equiv \int dr_1\, dr_2\, \rho(r_1, 0)\, \rho(r_2, t)\, \delta(r_1 - r_2)$$
$$= \sum_{i=1}^{N} \sum_{j=1}^{N} \delta\big(r_i(0) - r_j(t)\big). \tag{25}$$

The temporal evolution of the translational non-Gaussian parameter $\alpha_2(t)$ and four-point dynamic susceptibility $\chi_4(t)$ for a series of temperatures under study is shown in Supplementary Material (Figures S6 and S7).

In **Figure 6,** we show the time evolution of $\alpha_2(t)$ and $\chi_4(t)$ for the fast and slow modes in the bifurcated spectrum at T = 0.6 and 0.55. *We find that dynamics is heterogeneous in both the domains but much more so in the slow domains*. Further, as the temperature is lowered, the slow-mode population becomes more dynamically heterogeneous. What is particularly remarkable about Figure 6 is the large separation between the time scales of relaxation of the fast and slow peaks. While the time scales in the bifurcation differ only by a factor of 3-5, the time scales in the non-Gaussian and dynamic heterogeneity parameters differ by almost two orders of magnitude. This observation needs further investigation.

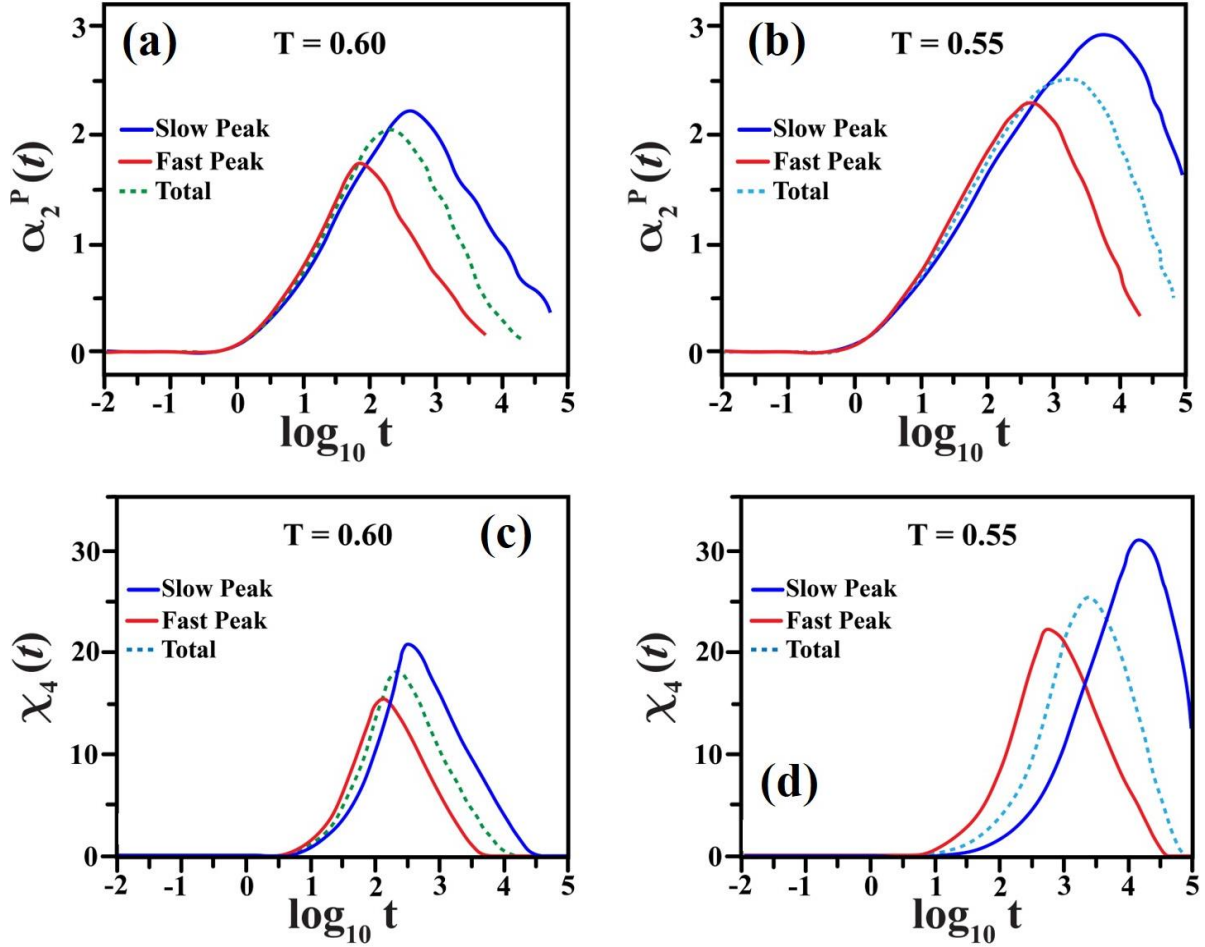

**Figure 6.** The temporal evolution of the translational non-Gaussian parameter $\alpha_2(t)$ of prolates corresponding to the population of slow and fast mode of the relaxation spectra at (a) T=0.60 and (b) T=0.55. In panels (c) and (d), we show the time evolution of the four-point dynamical susceptibility $\chi_4(t)$ for the population corresponding to the slow and fast modes at T=0.60 and (b) T=0.55, respectively. It is to be noted that a significantly higher extent of dynamic heterogeneity exists in the population of slow mode.

Since there is no phase separation (as verified by radial distribution function, see Figure S8), this result seems to suggest the existence of a mosaic-like structure. We point out two important points. First, the slow domain is characterized by less entropy but more enthalpic stabilization than the fast domain. We reach this conclusion by using the diffusion-entropy scaling relation,[108] which is semi-quantitatively reliable. Second, at low temperatures, a major source of relaxation could be the conversion of the slow domains to fast domains. The pathway of such relaxation is unclear. It is interesting to compare this behaviour with LDL-HDL conversion observed in supercooled liquid water.[109,110]

**Figures 7(a) and 7(b)** present the product of translational diffusion constant $D$ and the second-rank rotational relaxation time $\tau_2$ for prolates and oblates, respectively. The idea to plot $(D \times \tau_2)(T)$ is to test the coupling/decoupling between the translational and rotational motion of particles, akin to the Debye-Stokes-Einstein relation.[73] This approach parallels the assessment of coupling/decoupling between viscosity and diffusion in the Stokes-Einstein relation, which is a standard practice. In the normal liquid regime, this product is typically independent or weakly dependent on temperature. This product is a popular combination often employed to understand the emergence of dynamic heterogeneity. The basic idea is as follows. As dynamic heterogeneity becomes pronounced, the faster regions determine the average diffusion coefficient, while the slower regions determine the rotational correlation time. It has been argued that the rotational correlation time scales with viscosity while translational diffusion decouples from viscosity.

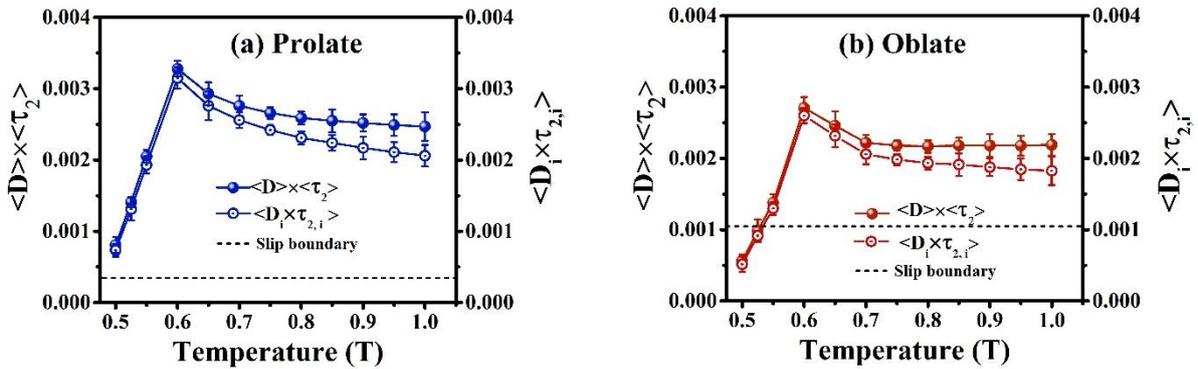

**Figure 7.** The plot of $\langle D \rangle \times \langle \tau_2 \rangle$ and $\langle D_i \times \tau_{2,i} \rangle$ as a function of temperature for (a) prolates and (b) oblates. The hydrodynamic prediction using slip boundary condition is shown by the dotted line. A significant deviation from the hydrodynamic prediction is observed.

As, in general, the product $\langle D \rangle \times \langle \tau_2 \rangle$ of the averages $D$ and $\tau_2$ is not identical to the average $\langle D_i \times \tau_{2,i} \rangle$, we further explore the temperature dependence of $\langle D_i \times \tau_{2,i} \rangle$. *In the present* *case, both the products exhibit a surprising maximum at a temperature higher than the estimated glass transition temperature.* For convenience, we term the temperature where the maximum is located as a crossover temperature, $T_{cr}$. As the temperature $T$ approaches this crossover temperature from above, the product first increases. This increase above the temperature-independent value at higher temperatures is well-known and often referred to as a signature of the emergence of dynamical heterogeneity. Much above $T_{cr}$, the product $D\tau_2$ becomes independent of temperature, which is expected in the normal regime of the liquid if it obeys the Debye-Stokes-Einstein relation. In the same figure, we showed the prediction of the hydrodynamic slip boundary condition for prolate and oblate particles.[111]

The rather sharp fall in the product below the crossover temperature $T_{cr}$ is puzzling and may be attributed to the *conversion of the liquid-like domains to the solid-like domains*, as the temperature is lowered below $T_{cr}$. As a result, $D$ decreases faster than the increase in $\tau_2$. This difference could be facilitated by the fact that rotation is still a local process, and our particles are not too different in shape from spheres with aspect ratios of 1.2 (prolate) and 0.8 (oblate). Thus, it is the diffusion coefficient, which is non-local, that undergoes sharp change as we lower the temperature below $T = 0.6$, which is close to the MCT critical temperature (from power-law fits).

What is more surprising is the sharpness of the peak in the product, and it is more difficult to rationalize. We note that the temperature variation of both $D$ and $\tau_2$ can be individually fitted to the Vogel-Fulcher-Tammann (VFT) form with the same glass transition temperature $T_g \sim 0.43$, which is substantially below the crossover temperature, $T_{cr} \sim 0.6$ (as shown in Figure 7). The fitting parameters are sensitive to lower values of temperature.

In theoretical studies, the existence of such crossover temperature has repeatedly been discussed.[10,49,67,68,112] The mode coupling theory predicts a divergence of viscosity and relaxation time at a temperature that is substantially above the true glass transition temperature. MCT, of course, does not provide a microscopic mechanism except through the growth of structural correlations among particles.

A physical explanation of a crossover temperature is obtained by analysis of the inherent structures. By calculating the average energy of the inherent structures obtained from molecular dynamics (MD) trajectories and analyzing the dynamics, one can locate a temperature where the dynamics change from continuous diffusion to a landscape-dominated regime.[15]

In an early work, Ediger et al. observed such a crossover in terms of the emergence of heterogeneity and provided compelling evidence in terms of the decoupling between translation and rotation by the mechanism we already discussed above.[81] However, the existence of a sharp peak in the product $D\tau_2$ has not been discussed earlier.

In the present case, we find that the crossover temperature is close to the initial appearance of the bifurcation in the relaxation spectrum. To provide a theoretical understanding of the observed bifurcation, we present a dynamic exchange model (DEM) discussed in detail in the Supplementary Material.

## D. Nature of the observed bifurcation: Johari-Goldstein or MCT?

In this section, we explore the potential relationship of the observed bifurcation with existing theoretical frameworks such as the Johari-Goldstein (JG) α-β bifurcation or that predicted by the Mode-Coupling Theory (MCT). For this purpose, it is imperative to clarify our use of terminology in this context. We called the bifurcation in orientational relaxation as Johari-Goldstein because the latter discussed it in the context of dielectric relaxation. Johari and Goldstein used aspherical dipolar molecules. The α-β bifurcation in the JG picture was so categorized by *borrowing ideas from polymer dynamics* where the β-relaxation exhibits Arrhenius temperature dependence and is attributed to side-chain motions. This β-relaxation is weakly coupled to the main branch of relaxation.

Such a picture is not expected to be valid here. We have small molecules with both translation and rotational degrees of freedom. Contrastingly, in the MCT framework, the initial beta relaxation stems from "local" relaxation occurring over relatively short timescales before transitioning into the characteristic staircase behaviour. This can be likened to density relaxation within a potential energy well, as envisioned in the inherent structure description. Strictly speaking, one can attribute this only to the early stages of a continuous density relaxation process that does not involve any barrier-crossing dynamics.

We have already observed in Eqs. (12-15) and Table 1, a fairly good agreement of our results with the important MCT prediction (Eq. 14). This actually provides a striking confirmation of the MCT.

In order to examine further the potential relationship of the observed bifurcation to that predicted by MCT, we compute the susceptibility functions from the self-intermediate correlators $\left(F_s^{P/O}\right)(k_{max},t)$ and $C_2^{P/O}(t)$. For this, first of all, we have calculated the derivative of $\left(F_s^{P/O}\right)(k_{max},t)$ and $C_2^{P/O}(t)$, followed by the Fourier transform. **Figures 8(a) and 8(b)** illustrate the imaginary part of the susceptibility functions derived from the Fourier transform of the first derivative of the correlators for prolates. We observe that at high temperatures, both cases exhibit a single peak; however, as the temperature decreases, we observe a bifurcation akin to the behaviour observed for $P(D,T)$ and $P(\tau_2,T)$. The observed bifurcation in susceptibility functions closely aligns with those reported by Kämmerer et al.,[49] further bolstering the consistency and reliability of our findings.

As discussed, the difference between the Johari-Goldstein and MCT α–β picture of the underlying bifurcation mechanism is indeed significant. *However, it is essential to note that the properties of the observed α-β bifurcation in our system may differ from both the JG β-peak and the MCT β-peak.* For instance, the location of the maximum and the critical behaviour of the peaks can vary. In addition, in our case, remnants of β-relaxation are observed in the extended ripple-like region in the tail of the probability distribution, as evident in both $P(D)$ and $P(\tau_2)$ at $T = 0.5$. This is more like the prediction of the JG picture.

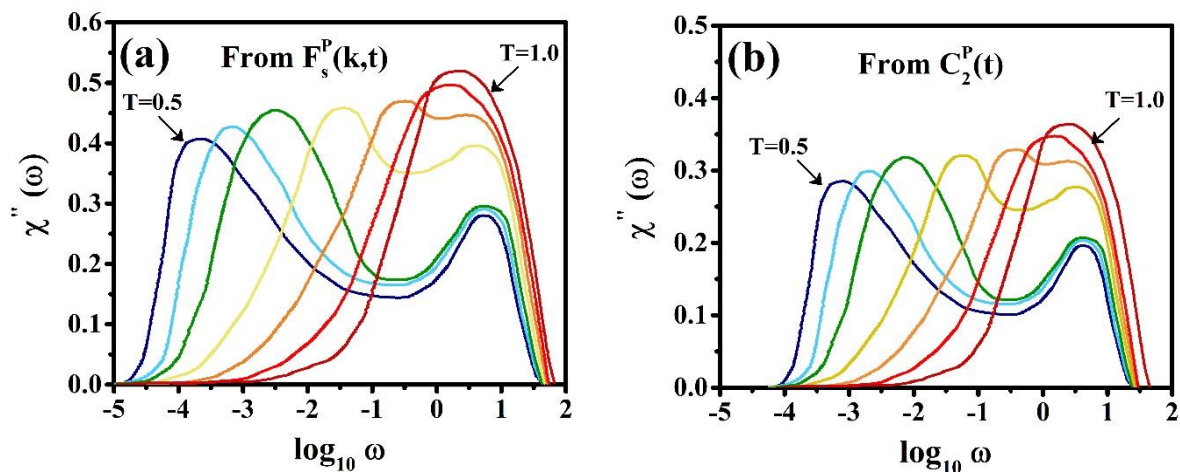

**Figure. 8. Frequency dependence of the imaginary part of the susceptibility functions derived from the Fourier transform of the first derivative of the correlators (a)** $\left(F_s^P\right)(k_{max}, t)$ **and (b)** $C_2^P(t)$**. It is to be noted that, in both cases, a single peak is observed at high temperatures; however, as the temperature is lowered, a bifurcation of the peak is observed. The colour code is same as that of Figure 2.**

### E. Jump dynamic transitions in the dynamics of mosaics

What really drives relaxation in deeply supercooled liquids has remained an open question. Inherent structure (IS) formulation advocates transition between different glassy minima in the potential energy landscape.[15,59] However, in the inherent structure formalism, the mechanism of such transitions is not addressed. The random first-order theory assumes that such relaxation occurs between different mobility states via a nucleation process.[10] The transition is akin to melting-freezing transitions driven by the entropy difference between the frozen (the glassy) and the mobile (the liquid) states. Such transitions are predicted to occur in regions that may contain a small number of particles ranging from a few tens to a few hundreds. In other words, these transitions are collective in nature (not single particle), involving mesoscopic regions rather than macroscopic ones.

In this subsection, we present a few microscopic characterizations that aid in understanding certain aspects of the glassy relaxation dynamics observed in this paper. Figures S9(a) and S9(b) demonstrate the presence of correlated translational and rotational jump motions of tagged single-particle prolate and oblate particles at a low temperature, $T=0.6$. Such single-particle jumps have been observed earlier in supercooled liquids, with varying interpretations derived from inherent structure hopping and single-file diffusion.[50] In Figure S9(c), we show the translational displacement trajectory of a randomly selected tagged prolate (say $N_P=752$), illustrating the jump motion at T=0.60. The same figure depicts the trajectories of the nearest neighbours of the randomly selected tagged prolate where correlated jump motion has been observed. However, it is not clear what role these correlated single-particle jump motions play in the overall relaxation dynamics of the system because the influence of such jumps appears to be small.

In the RFOT analysis, on the other hand, relaxation is assumed to occur through transitions in a region, accompanied by small length-scale correlated motions of the particles involved.[10] In this process, each particle moves only about the Lindemann length, typically one-tenth of the molecular diameter. The main idea is that to make a transition or significant movement in the free energy surface, many particles need to move, but each a small distance, as in the liquid-solid transition. It is possible that the above-mentioned large-scale jumps form an ensemble of relaxation processes that are too rare to make any significant impact.

To discern the occurrence or absence of such transitions within local regions and gain further insight into the heterogeneous dynamics near the glass transition (and to better comprehend the observed bifurcation), we carried out the following analysis. We divided our entire system into 64 cubic grids, each containing approximately 62-63 particles, and monitored the dynamical properties in these cubic boxes as a function of time. In order to analyze the evolution of diffusion constants and orientational relaxation times within specific spatial regions over time, we divided the trajectory into discrete time intervals, each spanning approximately $10^7$ steps. Within each interval, we computed both the diffusion constant and the orientational relaxation time. This approach allows us to identify spatial variations in dynamics and potential localized phenomena contributing to system behaviour. Such a strategy not only offers a comprehensive understanding of the dynamics of the system but also provides valuable insights into the role of local environments in governing macroscopic behaviour.

In **Figures 9(a) and 9(b)**, we show the short-to-intermediate time diffusion of a few grids as a function of time at T=0.6. In **Figures 9(c) and 9(d)**, we show the intermediate-time orientational relaxation times ($C_2(\tau_2) = e^{-1}$) of a few grids as a function of time at T=0.6. It is to be noted that while in many grids, there is no significant change in the diffusion constant (or in the orientational relaxation time), in some of the grids, we observe large-scale sudden changes in the diffusion coefficient and rotational correlation time. The jump magnitude in the time-averaged properties is large. *It appears to signify local first-order transitions between low and high-mobility domains.* Further, we have also observed a correlated change in the short-to-intermediate time diffusion coefficient and the orientational relaxation time. Similar results have been observed at T=0.55, as shown in the Supplementary Material (Figure S10); however, the frequency of these jumps is low.

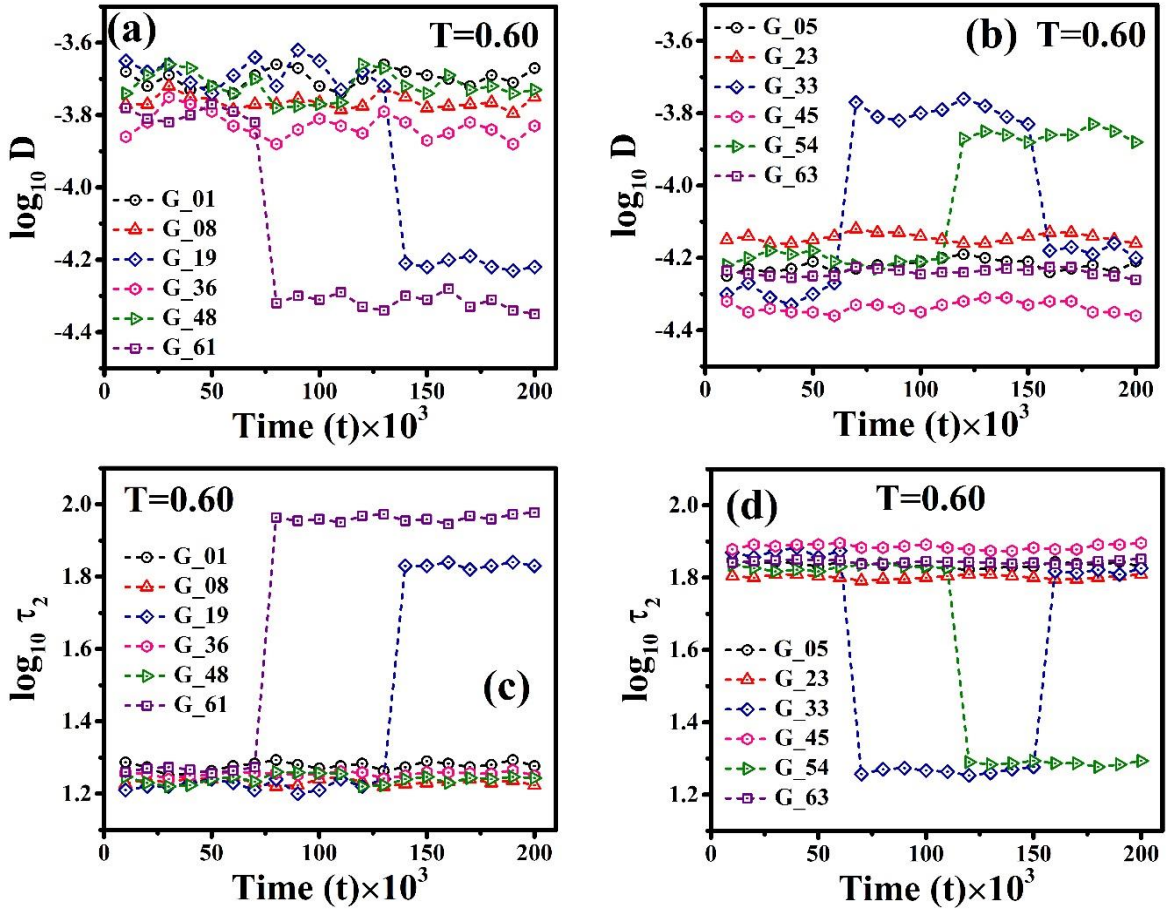

**Figure 9. (a)-(b) The variation of the short-to-intermediate time diffusion constant of particles in grids as a function of time at T=0.60. (c)-(d)The variation of the intermediate time orientational relaxation times ($C_2(t)$) of particles in grids as a function of time at T=0.60. We observe a sharp change in the diffusion constant of some grids, indicating liquid-to-solid-like and solid-to-liquid-like transitions. Further, a correlated change in the short-to-intermediate time diffusion and orientational relaxation time has been observed.**

Furthermore, we find that the initial values and the magnitude of the change in the transport properties due to the jumps depicted in Figure 9 (both in the short-to-intermediate time diffusion constant and the rotational relaxation time) correlate well with the values in the two peaks of the α-β bifurcation pictures presented above (Figure 5).

Although we have not been able to construct yet a complete waiting time distribution of these large amplitude jumps in the transport properties of our coarse-grained boxes, we have enough statistics to establish that these jumps can be considered rare in our simulation time scales, happen at a time scale of the order of hundreds of ns. That is, much longer than molecular time scales.

The next question arises: what prompts such jumps? Our analysis (still going on) points to the following picture. An upward jump in mobility, that is, the diffusion constant (and simultaneous downward jump in the rotational correlation time), is found to be accompanied by a density fluctuation, leading to a small but discernible decrease in its value. The density change is small as only one or two molecules are found to leave the box, followed by an increase in mobility (Figure S11). The reverse transition, that is, a decrease in mobility, is found to be accompanied by a small increase in density. This is similar to the changes observed in the HDL-LDL transitions, where a similar coarse-grained analysis was performed.[110] Such density changes are clearly accompanied by a significant entropy change. We further note that we count the inclusion of particles in a box by monitoring its centre. Therefore, the particle can indeed move a small distance to be considered out of a given box. We mention this analysis because these changes, however small, are observed only during or just before the jumps described in Figure 9. A lot more simulations and analyses are required to establish the detailed origin, especially correlations between neighbouring boxes.

A further point that should be made clear about the time scales observed here, because all results have been presented in the reduced units. Yet, we sometimes need absolute numbers to get additional insight into the relaxation dynamics. We can convert our numbers presented here to real argon units easily. In the argon unit, the translational diffusion constant decreases from $10^{-5} cm^2 s^{-1}$ to about $10^{-10} cm^2 s^{-1}$. That is, we observe a five-order-of-magnitude decrease from the liquid at $T = 1.0$ to the glassy liquid at $T = 0.5$. Similarly, the rotational correlation time decreases from ps to us. It is thus clear that the present simulations capture the initial part of the slowdown toward glass transition.

## IV. CONCLUDING REMARKS

Unresolved or partly solved questions in the dynamics of supercooled (or glassy) liquids pertain to the relationship between the α-β bifurcation temperature $T_B$, popularized by Johari and Goldstein in the late 1960s, and the crossover temperature $T_{cr}$ observed in simulations have remained the subject of much interest. Such a relationship has been predicted by all the major theories developed after Johari-Goldstein's observations. MCT, inherent structure analysis, and RFOT all predict different signatures for the existence of a crossover temperature above the glass transition temperature. We note that the dielectric relaxation measured by Johari and Goldstein involved primarily the *orientational motion of molecular dipoles.* The molecules involved were anisotropic. However, most computer simulations and theories employ spherical-shaped molecules, like in the popular Kob-Andersen model of glass forming binary mixture.[6,7] A recent work by Cicerone et al. reports the evidence of such a bifurcation at a much shorter ps time scale.[38]

We note that such a bifurcation, as reported in **Figure 5**, can also be expected in spherical models in the short-to-intermediate time diffusion coefficient. *Therefore, anisotropy in shape might not introduce any substantially new feature, except allow one to address rotational relaxation and the translation-rotation coupling*. The present model may be considered a generalization of the well-known Kob-Andersen binary mixture that consists of spheres of unequal size and disparate interaction. The Kob-Andersen model is known to reproduce several aspects of glass transition behaviour. It reproduces the divergence of viscosity and captures the non-exponential structural relaxation measured by neutron scattering experiments, but having no orientation, cannot really address the Johari-Goldstein mode(s).

We first demonstrate that the present model with anisotropy in shape reproduces most of the experimentally observed thermodynamic changes as the temperature approaches the glass transition temperature. *For example, we find a sharp rise in the specific heat when a* *rapidly cooled system is heated again.* The thermal expansion coefficient also undergoes a sharp change. Transport properties provide a glass transition temperature of around 0.43. Both prolates and oblates give the same glass transition temperature.

We observed several potentially interesting results in this study. First is the observation of α−β bifurcation in the relaxation of the orientation and density relaxation, as shown in **Figure 5**. What is remarkable is that the appearance of the bifurcation nearly coincides with the crossover temperature captured in the product $D\tau_2$, also plotted in **Figure 7.** Thus, we could conclude that $T_{cr} \sim T_B$, a conclusion that agrees with Stevenson-Wolynes' theory.[37] Another important result is that the bifurcations take place in the ps to ns time scales. This agrees well with the recent report of such bifurcation by Cicerone et al.[38]

Relatively sharp non-Arrhenius rise of viscosity or relaxation time in the Angell fragility plot is often attributed to the entropy crisis scenario, as in Adam-Gibbs or RFOT theories. The presence of bifurcation in the relaxation spectrum found in the present study suggests that the bend in the relaxation time (or viscosity) versus $1/T$ in the Angell plot arises partly from the conversion of the liquid-like domains to solid-like domains. This is also clearly present in the mosaic picture, and in that sense, it is not fully surprising.

There seems to exist an interesting analogy with the HDL-LDL transition in supercooled liquid water.[110] As the temperature is lowered below 273K, the fraction of LDL increases. Theoretical study again reveals the appearance of a mosaic-like structure that alternates between the two states of liquid water. As the temperature is lowered further below 230K, only the LDL domains occupy all regions.

A particularly important outcome of the present study is that both the non-Gaussian order parameter $\alpha_2(t)$ and the dynamic heterogeneity parameter $\chi_4(t)$ clearly demonstrate the differing dynamic state of the two peaks in the relaxation spectrum. Furthermore, the peak time of these quantities differs by 2-3 orders of magnitude. Thus, fluctuations in the liquid-like domains are shorter-lived than those in the solid-like domains. However, what is interesting to note is that this difference is much larger than a factor of 3-5 difference we observe in the value from the relaxation spectrum itself.

Neither the present model nor the Kob-Andersen models admit to a crystalline state, and this poses a problem because most experimental results are discussed in terms of the melting temperature of the stable crystalline phase, and one often finds that $T_g \sim (2/3)\ T_m$, where $T_m$ is the melting temperature. In the present case, we again find $T_g \sim (2/3)\ T_{cr}$. *A tantalizing possibility is that $T_{cr}$ is still indirectly correlated with $T_m$,* in the case of real liquids that admit a crystalline state. This proposition needs to be verified. In the mosaic picture and the RFOT theory, the low entropy, slow, solid-like domains are expected to be better packed and more ordered if a crystalline state is allowed to exist.

*To conclude, we have introduced a new class of models of glass-forming liquids. The model admits to orientational motion.* We find a Johari-Goldstein-type bifurcation in the rotational relaxation at a temperature $T_B$. This temperature is close to the crossover temperature observed in the product $D\tau_2$. While the sharp fall in the product $D\tau_2$ below $T_{cr}$ reflects the markedly different dynamics probed by rotation and translation, the low value of the product near the glass transition temperature seems to suggest the freezing out of the remaining liquid-like regions or domains that take place between a reduced temperature of 0.60 and 0.43.

## SUPPLEMENTARY MATERIAL

See **Supplementary Material (SM)** for the following: (i) temperature dependence of specific volume and isothermal compressibility, (ii) fitting of self-intermediate scattering function to a power-law form, (iii) temporal evolution of mean-square displacement (MSD) of prolates and oblates, (iv) bifurcation spectrum of oblates, (v) dynamic heterogeneity and non-Gaussian parameters. (vi) partial radial distribution functions (RDFs) of prolates and oblates, (vii) dynamic exchange model (DEM) to provide theoretical analysis of bifurcation, (viii) correlated translational and rotational jump motions and (ix) jump dynamic transitions in the dynamics of mosaics at $T$ =0.55.

ACKNOWLEDGEMENTS

We thank Dr. Tuhin Samanta for collaboration at the incipient stage, Dr. Sayantan Mondal, Dr. Saumyak Mukherjee and Ms. Sangita Modal for discussions. B.B. thanks the SERB-DST, India, for than India National Science Chair (NSC) Professorship and for providing partial financial support. S.K. thanks the Council of Scientific and Industrial Research (CSIR) for a research fellowship.

AUTHOR DECLARATIONS

**Conflict of Interest**

The authors have no conflicts to disclose.

DATA AVAILABILITY STATEMENT

The data that supports the findings of this study are available within the article and its Supplementary Material.

# Supplementary Material

# Glassy Dynamics in a Molecular Liquid

**Shubham Kumar, Sarmistha Sarkar and Biman Bagchi***

*Solid State and Structural Chemistry Unit, Indian Institute of Science, Bangalore 560012, India.*

*****Email: bbagchi@iisc.ac.in; profbiman@gmail.com**

**SUPPLEMENTARY MATERIAL TEXT**

In the following, we present several results that supplement the results discussed in the main text. We briefly elaborate on some of the results.

## (i) Thermodynamic properties

In the main text, we have discussed the temperature variation of the thermal expansion coefficient $(\alpha)$ and the specific heat at constant pressure $(C_P)$ across the glass transition temperature. In **Figure S1**, we show the variation of the specific volume $(V_{sp})$ as a function of temperature from molecular dynamics (MD) simulations in the NPT ensemble as well as for cooling-reheating cycles (with three different constant cooling/heating rates). Near the glass transition, a change in the slope of specific volume is observed, which is consistent with the earlier studies with spherical atom models.[1] The glass transition temperature $(T_g)$ is predicted by the intersection of the curves (when extrapolated) on both sides of the transformation range. The temperature-dependent variation during the cooling/heating is found to be sensitive to the rate of the processes. However, the dependence of $T_g$ o`n the cooling/heating rate is weak.

Along with the thermal expansion coefficient $(\alpha)$ and the specific heat at constant pressure $(C_P)$, isothermal compressibility is required for the estimation of the Prigogine-Defay Ratio (PDR).[2] The isothermal compressibility in the NPT (isothermal-isobaric) ensemble is defined by the fluctuation in the specific volume as follows,[3]

$$\kappa = -\frac{1}{V_{sp}}\left(\frac{\partial V_{sp}}{\partial P}\right)_T = \frac{1}{V_{sp}k_B T}\sigma^2_{V_{sp}} \tag{S1}$$

Here, $\sigma^2_{V_{sp}} = \left(\langle V_{sp}^2 \rangle - \langle V_{sp} \rangle^2\right)$.

**Figure S2** depicts the variation of isothermal compressibility (obtained from MD simulations in the NPT ensemble) as a function of temperature across the glass transition region. Similar to the thermal expansion coefficient $(\alpha)$ and heat capacity, it also shows a sharp change near the glass transition.

We obtain a Prigogine ratio of 2.7 for our model system.

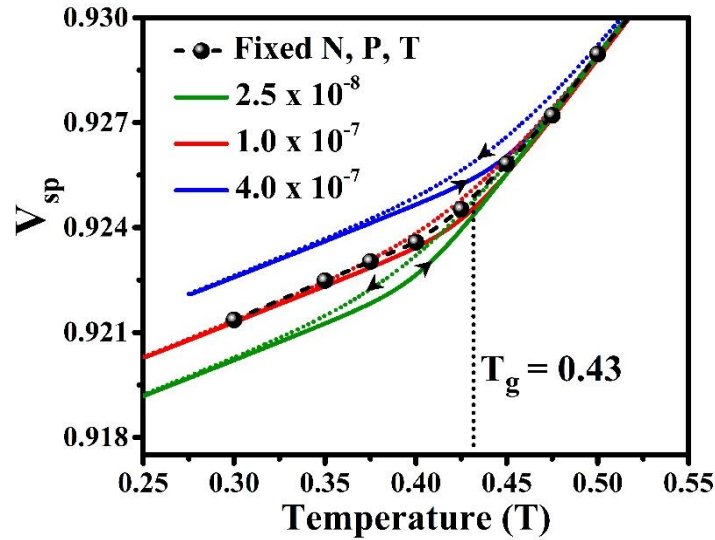

**Figure S1. The temperature dependence of specific volume. The results are obtained by MD simulations in the NPT ensemble (black spheres with dashed line), isobaric cooling (dotted lines), and isobaric heating (solid lines) of the system for three different constant cooling/heating rates. A change in the slope of the specific volume shows a signature of glass transition. The glass transition temperature further depends on the rate of heating/cooling. From MD simulations in NPT ensemble, the glass transition temperature obtained is $T_g \sim 0.43$.**

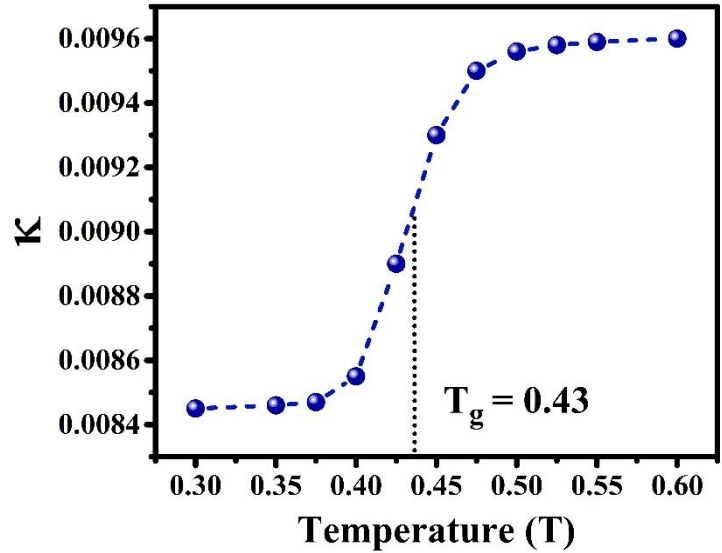

**Figure S2.** The variation of isothermal compressibility with temperature across the glass transition region. The results are obtained by MD simulations in the isothermal-isobaric (NPT) ensemble. A sharp change in the isothermal compressibility shows a signature of glass transition, with the glass transition temperature at $T_g \sim 0.43$. The dotted lines are drawn as a guide to the eyes.

## (ii) Fit of self-intermediate scattering function to power law forms

**Figure S3** shows the time variation of the self-intermediate scattering function in a semi-log plot. This figure shows the three regimes familiar in the glass transition studies: (i) at short times ballistic motion of particles, followed by the transient trapping in cages (β-relaxation), (ii) very slow decay (plateau-like behaviour) at intermediate times, and (iii) at sufficiently long times, the fall from the plateau which is known as the α-relaxation.[4] We also find the two power-law decays separated by a plateau. The results here depict the behaviour just above the power law prediction of the glass transition temperature, which, as mentioned in the main, is found by fitting to be 0.54.

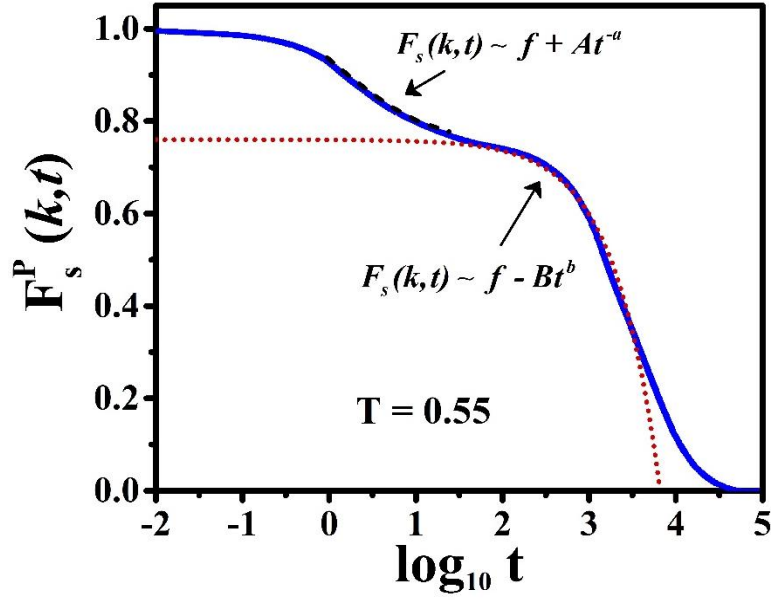

**Figure S3.** Time dependence of self-intermediate scattering function, $F_s(k,t)$ of prolate at T =0.55. The decay to the plateau and from the plateau region is governed by a power law. The fittings are shown by the dotted lines.

In **Table S1.** we provide the power law fitting parameters of self-intermediate scattering functions at T = 0.55.

**Table S1.** Parameters obtained by the power law fitting of decay to the plateau and from the plateau region of self-intermediate scattering functions at T=0.55.

|         | $f$  | $A$  | $a$  | $B$              | $b$  |
|---------|------|------|------|------------------|------|
| Prolate | 0.76 | 0.17 | 0.37 | $5.6\times10^{-4}$ | 0.82 |
| Oblate  | 0.79 | 0.13 | 0.39 | $6.2\times10^{-4}$ | 0.81 |

### (iii) Mean-square displacement of prolates and oblates

In **Figure S4**, we present the time dependence of the mean square displacements (MSD) of prolates and oblates for different temperatures employed in the present study. At short times, the motion is ballistic, and at high temperatures, the ballistic motion is rather quickly followed by the diffusive motion. The diffusion constants are obtained by the linear fitting of mean square displacement versus time curves (by avoiding the initial ballistic region). The situation changes at low temperatures. Figure S4 shows a dramatic slowdown in the mean square displacement below temperature 0.6, *implying that time scales of relaxation stretch rapidly as temperatures fall below 0.6, which is close to our cross-over temperature.*

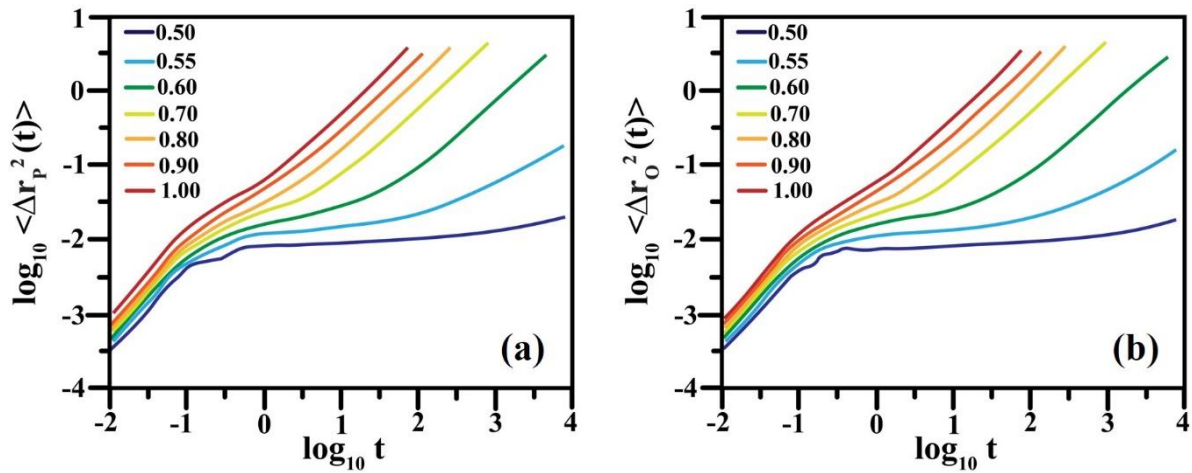

**Figure. S4. Time dependence of the mean square displacement (log-log plot) of the (a) prolate and (b) oblate for a range of temperatures starting from low (T = 0.50) to high temperature (T = 1.0). It is to be noted that at short times, the motion is ballistic; however, in long times, diffusive motion is observed.**

### (iv) Bifurcation of the distribution of diffusion coefficient of oblates

In the main text, we discussed the distribution function of the prolates, which is the main constituent of the binary mixture. In **Figure S5**, we plot the distribution of the short-to-intermediate time diffusion constant of oblates. As expected, oblates also show the appearance of the bifurcation above the glass transition temperature. The bifurcation becomes quite pronounced near $T = 0.6$, where the peak of the product between the diffusion constant and rotational correlation time is placed.

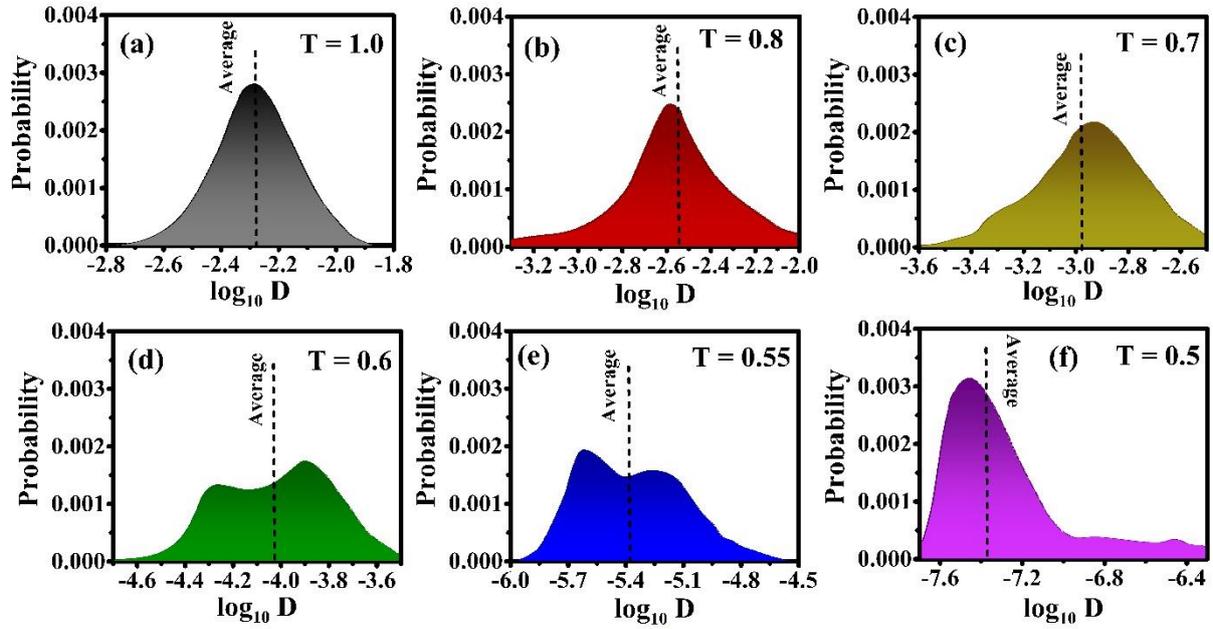

**Figure S5.** The distribution of the short-to-intermediate time diffusion coefficient of oblates at (a) T = 1.0, (b) T=0.8, (c) T=0.7, (d) T=0.6, (e) T=0.55, and (f) T=0.5. At a higher temperature, there is only one peak; however, as the temperature is lowered, a bimodal distribution is observed.

## (v) Dynamic heterogeneity and non-Gaussian parameters

The emergence of dynamic heterogeneity is a common signature of glassy liquids. This is quantified mainly by two functions, the non-Gaussian parameter (NGP), denoted by $\alpha_2(t)$, and a four-point correlation function, $\chi_4(t)$.

**Figure S6** depicts the temporal evolution of the translational non-Gaussian parameter (TNGP) for a series of temperatures. It is to be noted that the behaviour looks similar for both the components in the binary mixture. As the temperature is lowered, the peak gradually shifts to higher times, varying by over four orders of magnitude.

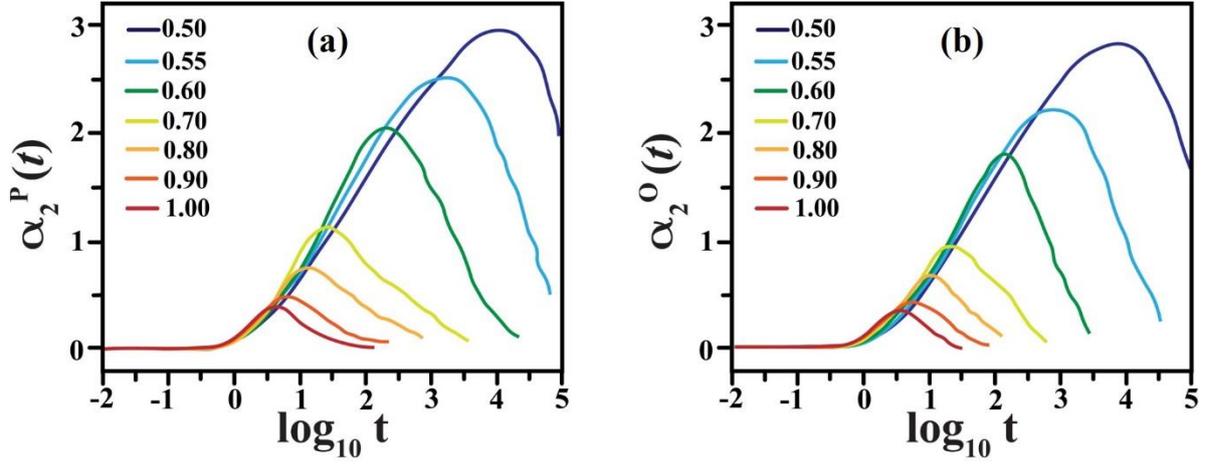

**Figure S6.** Temporal evolution of the translational non-Gaussian parameter (TNGP) of (a) prolates and (b) oblates for a series of temperatures investigated.

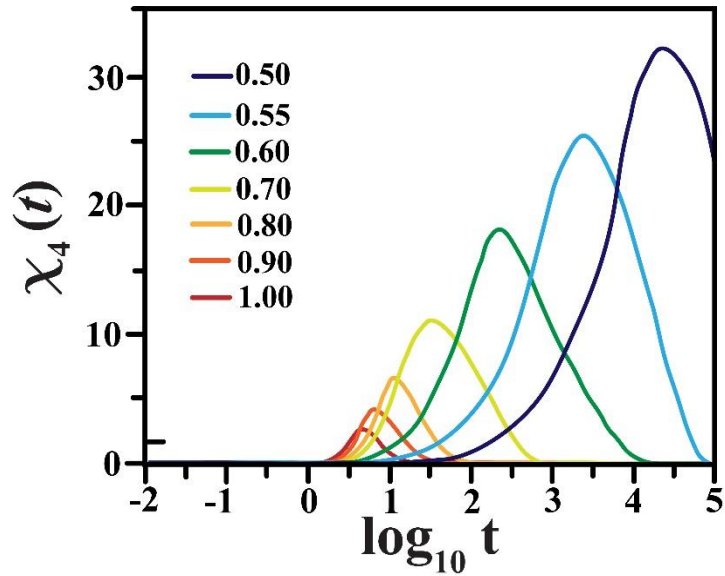

**Figure S7.** Time dependence of the four-point dynamic susceptibility $\chi_4(t)$ (a) prolates and (b) oblates for a series of temperatures investigated. The system becomes more heterogenous as we move towards the glass transition.

Besides the translational non-Gaussian parameter, the growth of dynamic heterogeneity in the supercooled state can be measured by the four-point dynamical susceptibility. In **Figure S7**, we show the time evolution of $\chi_4(t)$ for a series of temperatures. While both the four-point dynamic susceptibility and the translational non-Gaussian parameter signify that the system becomes more heterogeneous as we move toward the glass transition, the dynamic heterogeneity parameter exhibits sharper features.

## (vi) Partial radial distribution functions of prolates and oblates

In **Figure S8,** we present three partial radial distribution functions: (i) prolate-prolate, (ii) oblate-oblate, and (iii) prolate-oblate. These partial radial distribution functions describe local arrangements among the particles. Since the density of the respective species has been factored out by dividing by densities in the definition of the partial functions, these figures indeed represent the relative local preponderance of the two particles. We find that at both temperatures, *the oblate-prolate order is most pronounced*, followed by prolate-prolate and oblate-oblate. This arises from two reasons. First, we have introduced a large prolate-oblate attractive interaction, as in the Kob-Andersen model potential. Second, a possible better packing arrangement between prolate and oblate. This aspect is not too clear and needs further investigation. These features are expected to be important because they introduce frustration in the system and prevent crystallization, as in the Kob-Andersen model of a glass-forming binary mixture where the two spherical species A and B have different sizes and also attract each other more strongly than A-A and B-B interactions.

**Figure S8(b)** produces the bifurcation of the second peak, which is familiar in glass-forming liquids. This bifurcation originates in multiple possibilities of arrangements at the second nearest neighbour distance. In the hard sphere liquid, which is a good glass former, the split second peak is due to the difference between fcc and hcp lattice arrangements.

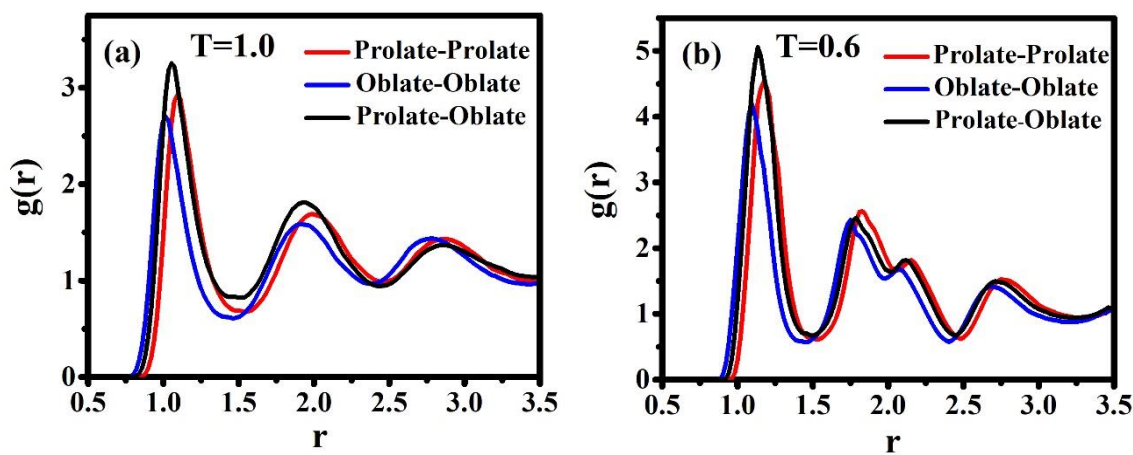

**Figure S8.** The plot of the partial radial distribution function of (a) the prolate-oblate binary mixture liquid at (a) high-temperature, T = 1.0 and (b) low-temperature, T = 0.6; at reduced pressure P = 30. At low temperatures, the characteristic of glass transition is indicated by the splitting of the second peaks.

## (vii) Theoretical analysis of bifurcation: Dynamic exchange model (DEM)

The observed rotational and translational dynamics are a complex mix of local dynamics and the non-equilibrium nature of the system that drives changes on mesoscopic length scales, such as the ones envisaged in the random first-order changes assumed in RFOT theory. Thus, when the surrounding liquid of a tagged rotating molecule undergoes a change to a solid-like domain, then the measured dynamics is a combination of both processes. On the other hand, if a solid-like domain undergoes a transition to a liquid-like domain, then a caged tagged molecule can start rotating. Thus, the emergence of a bimodal relaxation pattern can be thought of as a result of several factors, prominent among them are the liquid-to-solid and solid-to-liquid transitions.

The transitions are rare and can be considered as a Poisson process with different time constants for the liquid-like and solid-like domains. As the temperature is lowered, the fraction of solid-like domain increases, but as the system is in a non-equilibrium (or quasi-equilibrium) state, the use of detailed balance is questionable, and that limits our ability to reduce the problem.

Here, we present a simple theoretical analysis to address the dynamical events and the time scales involved in the bifurcation of relaxation obtained in Figure 5. We first assume a mosaic picture where the system is naturally divided into small liquid-like and solid-like domains, as envisaged in the RFOT theory. We next assume a stochastic dynamical exchange model where a liquid-like domain gets transformed to a solid-like domain and vice-versa, a scenario akin to freezing and melting processes.

We next define $P_L(r, \Omega_L, t)$, which denotes the joint probability of a molecule to reside at position $r$ in the liquid-like region and have an orientation, $\Omega$. Similarly, we define $P_S(r, \Omega_S, t)$ in the same fashion for solid-like domains. The following coupled equations of motion then follow,

$$\frac{\partial}{\partial t} P_L(r, \Omega_L, t) = D_T^L \nabla^2 P(r, \Omega, t) + D_R^L \nabla_\Omega^2 P_L(r, \Omega_L, t) - k_{LS} P_L(r, \Omega_L, t) + k_{SL} P_S(r, \Omega_S, t) \quad (S1)$$

$$\frac{\partial}{\partial t} P_S(r, \Omega_s, t) = D_T^S \nabla^2 P(r, \Omega, t) + D_R^S \nabla_\Omega^2 P_S(r, \Omega_S, t) - k_{SL} P_S(r, \Omega_S, t) + k_{LS} P_L(r, \Omega_L, t) \quad (S2)$$

where, $D_T^i$ and $D_R^i$ denotes the translational and rotational diffusion in the $i^{th}$ phase, $k_{LS}$ and $k_{SL}$ are the rates of transition from liquid to solid and vice versa. Thus, we have to deal with four rate (or time) constants. In the next step, we expand the density in the spherical harmonics,

$$P(r, \Omega, t) = \sum a_{lm}(r, t) Y_{lm}(\Omega), \quad (S3)$$

to obtain the following two coupled equations,

$$\frac{\partial}{\partial t} a_{lm}^L(k, t) = -[D_L k^2 + l(l+1) D_R^L] a_{lm}^L(k, t) - k_{LS} a_{lm}^L(k, t) + k_{SL} a_{lm}^S(k, t), \quad (S4)$$

$$\frac{\partial}{\partial t} a_{lm}^S(k, t) = -[D_S k^2 + l(l+1) D_R^S] a_{lm}^S(k, t) - k_{SL} a_{lm}^S(k, t) + k_{LS} a_{lm}^L(k, t). \quad (S5)$$

We now define two time constants:

$$\frac{1}{\tau_S(k)} = D_T^S k^2 + l(l+1)D_R^S$$
$$\frac{1}{\tau_L(k)} = D_T^L k^2 + l(l+1)D_R^L \quad \text{(S6)}$$

We now make the simplifying assumption that relaxation in the solid state is extremely slow, $D_R^S = 0$ and $D_T = 0$, so that the equation of motion for density relaxation in the solid state is given by the exchange dynamics only (this simplifying assumption can be relaxed easily). We then follow the standard procedure to obtain,

$$\frac{\partial^2}{\partial t^2} a_{lm}^L(k,t) = -[D_T k^2 + l(l+1)D_R^L]\dot{a}_{lm}^L(k,t) - k_{LS}\dot{a}_{lm}^L(k,t) + k_{SL}\dot{a}_{lm}^S(k,t)$$
$$= -l(l+1)D_R^L \dot{a}_{lm}^L(t) - k_{LS}\dot{a}_{lm}^L(t) + k_{SL}\{-k_{SL} a_{lm}^S(t) + k_{LS} a_{lm}^L(t)\}. \quad \text{(S7)}$$

In the rest of the analysis, we shall focus on dielectric relaxation only, which means that we need only the $k = 0$ component of $a_{lm}(k,t)$.

It is now simple and straightforward by standard procedure to eliminate the solid density term to obtain the following quadratic equation,

$$\frac{\partial^2}{\partial t^2} a_{lm}^L(t) + \left[l(l+1)D_R^L + k_{LS} + k_{SL}\right]\dot{a}_{lm}^L(t) + D_R^L l(l+1) k_{SL} a_{lm}^L(t) = 0 \quad \text{(S8)}$$

This simple quadratic equation with constant coefficients is solved to obtain two time scales,

$$k_{\pm} = -B \pm \sqrt{B^2 - 4C} \quad \text{(S9)}$$

where,

$$B = \left[l(l+1)D_R^L + k_{LS} + k_{SL}\right]$$
$$C = l(l+1)D_R^L k_{SL}$$
(S10)

According to this model, the two rate constants would represent the two peaks in the relaxation spectrum (Figure 5).

Next, it may be safe to assume that $k_{LS} \gg k_{SL}$. Thus, the rate constants are given by

$$k_f \equiv k_- \approx -2B$$
$$k_s \equiv k_+ \simeq -2\frac{C}{B}$$
(S11)

A simple analysis shows that in the limit of extreme slow down, *the two rate constants are given by $l(l+1)D_R + k_{LS}$, and just $k_{SL}$*. The above simple analysis serves to explain the occurrence of the bifurcation in the relaxation spectrum. At high temperatures, there is no exchange mechanism, and the dielectric relaxation time shall be given by *$1/2D_R$*.

One can easily extend the above analysis to include translational diffusion. The standard procedure gives riser to wavenumber dependent dynamical functions *$a_{lm}(k,t)$*, where $k$ is the wave number. The resulting constant analysis gives rise to an additive term of the form *$l(l+1)D_R + D_T k^2 + k_{LS}$*. The other rate remains invariant. The preceding analysis is, of course, valid at a somewhat longer length scale, or the value of the wavenumber is significantly lower than the peak value of the structure factor.

The above analysis, however, gets modified because of the exchange of a region between solid-like and liquid-like domains. Let us consider that we have exchanges that are of time scales comparable to rotational relaxation and/or translational motion, measured by dynamic structure factor. Then, we may obtain an average over the relaxation in liquid-like and solid-like domains. This aspect certainly serves to explain the proximity of the two peaks in the relaxation spectrum. At lower temperatures, the contribution of the liquid-like domain diminishes rapidly.

Another interesting conclusion from the above analysis is that it helps in understanding the sharp peak in the product $D\tau_2$ shown in Figure 7. While the average mean square displacement (MSD) of particles is expected to be dominated by the fast liquid-like domains, the measured rotational time correlation function is expected to be dominated by the slow domains. *This explains the perplexing observation: why a probe molecule appears to translate further during its correlation time. Actually, in this experimentation, one is led to compare between two different molecules – one is translating in the fast domain and the other rotating in the slow domain.* The appearance of the sharp peak in the product $D\tau_2$ can now be understood as the temperature where a large fraction of liquid-like domains transforms to solid-like domains, as also seen in Figure 9. Thus, the contribution of the fast-moving particles in liquid-like domains disappears. Rotational diffusion, already dominated by the solid-like domains, is relatively less affected. Also relevant is the fact that our oblates and prolates are not drastically different from the spherical shape, so they might not get retarded by a significant degree.

The presence of the bifurcation in the relaxation spectrum seems to suggest the absence of any large growing correlation length, at least beyond a few molecular diameters. There is an interesting analogy with spectroscopy. When we study vibrational relaxation or Raman spectrum from a system consisting of two species or two energy levels, we obtain two peaks only when the exchange rate is slower than the lifetime of the states. In the presence of fast exchange, the two peaks merge to give rise to a single peak. This is called motional narrowing, a well-known phenomenon in vibrational and NMR spectroscopy.

## (viii) Correlated translational and rotational jump motions

**Figures S9(a) and S9(b)** demonstrate correlated translational and rotational jump motions of a tagged prolate and oblate particles at a low temperature of $T$=0.60. While such correlated jump motions are not unexpected and have been reported earlier, a few features stand out. The rotational jumps are often in the range of 130-140 degrees. That is, the rotational jumps are not the full 180 degrees allowed by symmetry. Second, the translational jumps are typically in the range of 0.75-1.0 molecular diameters. Such jumps have been observed by many in other systems like the Kob-Andersen model, where, however, rotational jumps could not be studied.

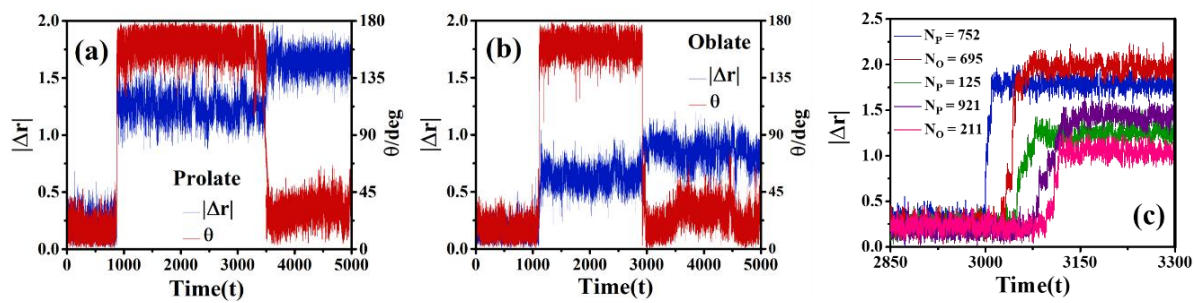

**Figure S9.** The coupled translational and rotational jump from the single particle trajectories for selected ellipsoids (a) prolate and (b) oblate over a time window at temperature T=0.60. (c) The magnitude of the translational displacement vector of a tagged prolate ($N_P$=752) and its nearest neighbours over a short time window illustrating correlated jump motion at T=0.60.

In **Figure S9(c),** we show the translational displacement trajectory of a randomly selected tagged prolate (say $N_P$=752), illustrating the jump motion at T=0.60. The same figure depicts the trajectories of the nearest neighbours of the randomly selected tagged prolate where correlated jump motion has been observed.

## (ix) Jump dynamic transitions in the dynamics of mosaics

In **Figures S10(a) and S10(b)**, we show the short-to-intermediate time diffusion of a few grids as a function of time at T=0.55. In **Figures S10(c) and S10(d)**, we show the intermediate-time orientational relaxation times ($C_2(\tau_2) = e^{-1}$) of a few grids as a function of time at T=0.55.

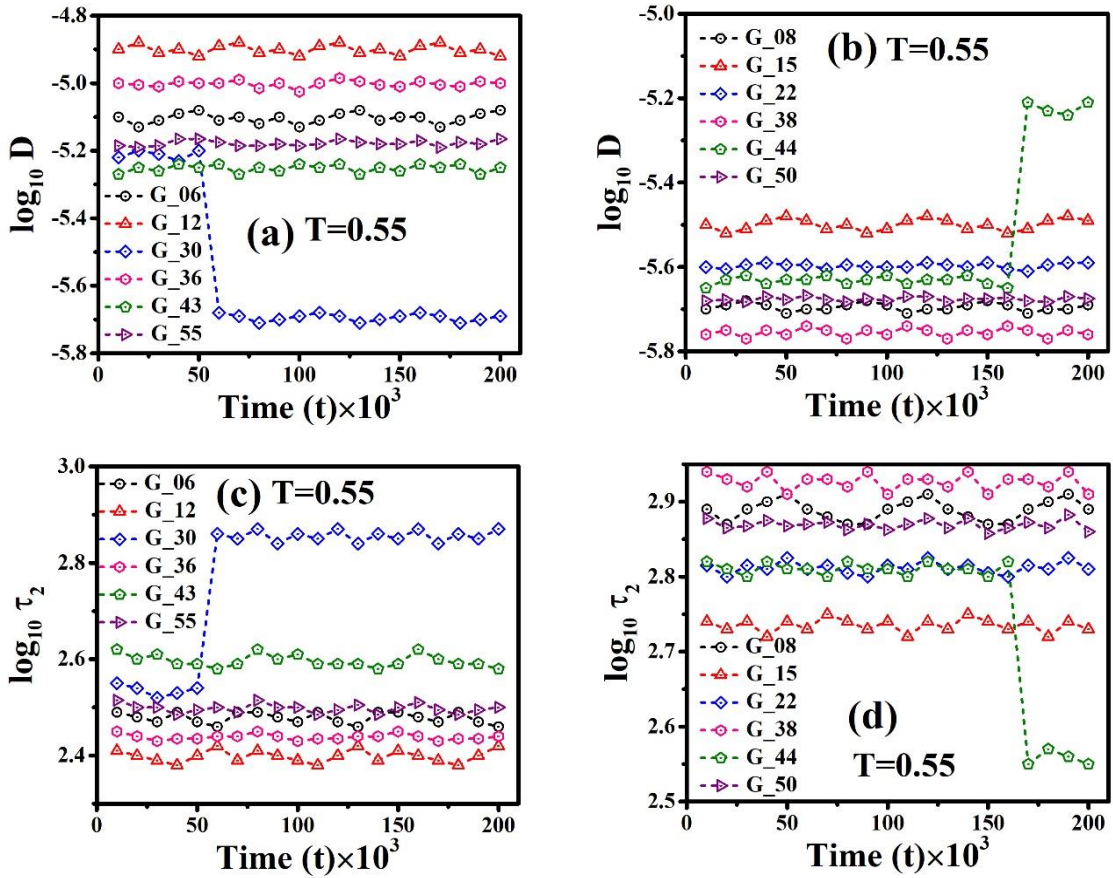

**Figure S10.** (a)-(b) The variation of the short-to-intermediate time diffusion constant of particles in grids as a function of time at T=0.55. (c)-(d)The variation of the intermediate time orientational relaxation times ($C_2(t)$) of particles in grids as a function of time at T=0.55. We observe a sharp change in the diffusion constant of some grids, indicating liquid-to-solid and solid-to-liquid like transitions. Further, a correlated change in the short-time diffusion and orientational relaxation time has been observed.

**Figure S11(a)** depicts the short-to-intermediate time diffusion constant of selected grids (say G-28) as a function of time at T=0.6. During the large-scale jump in the diffusion constant, two molecules are found to leave the box, followed by a small density change in the box, as shown in **Figure S11(b).**

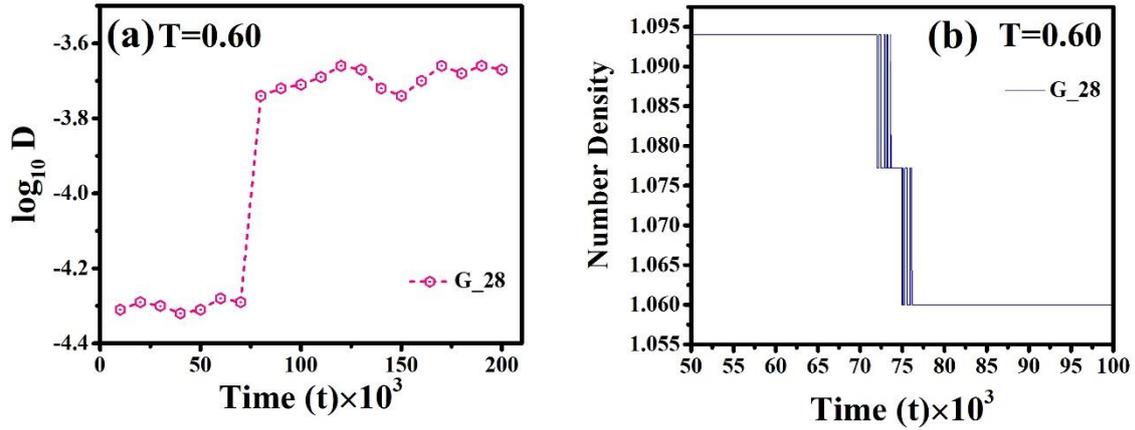

**Figure S11. The variation of the short-to-intermediate time diffusion constant of particles in a particular grid (G_28) as a function of time at T=0.6. (b) The density change in the box during the large-scale jump.**